\documentclass{aa}  

\usepackage{graphicx}
\usepackage{txfonts}
\usepackage{lipsum}
\usepackage{subcaption}         % necessary for continued figures, example in section 3
\usepackage{lscape}             % to rotate a single page table, example in appendix.
\usepackage{placeins}           % useful with \FloatBarrier, to keep 
\usepackage{blindtext}
\usepackage{tikz}
\usepackage{subfiles}
\usepackage{array}
\usepackage{hhline}
\usepackage{xcolor}
\begin{document}

   \title{High resolution ALMA observations of H$_2$S in LIRGS}

   \subtitle{Dense gas and shocks in outflows and CNDs}

   \author{M. T.Sato
          \inst{1}
          \and
          S. Aalto\inst{1}
          \and
          S. K\"{o}nig\inst{2}
          \and
          K. Kohno\inst{3}
          % \and
          % A. Fuente\inst{1}
          \and
          S. Viti\inst{4,5,6}
          \and
          M. Gorski\inst{7}
          \and
          F. Combes\inst{8}
          \and
          S. Garc\'{i}a-Burillo\inst{9}
          \and
          N. Harada\inst{10}
          \and
          P. van der Werf\inst{4}
          % \and
          % T. Izumi\inst{1}
          % \and
          % Y. Nishimura\inst{1}
          % \and
          % K. Sakamoto\inst{1}
          % \and
          % K. Alatalo\inst{1}
          \and
          J. Otter\inst{11}
          \and
          S. Muller\inst{1}
          \and
          Y. Nishimura\inst{3,}
          \and
          J. S. Gallagher\inst{12}
          \and
          A. S. Evans\inst{13,14}
          \and
          K. M. Dasyra\inst{15}
          \and
          J. K. Kotilainen\inst{16}
          }

   \institute{Department of Space, Earth \& Environment, Chalmers University of Technology, SE-412 96 Gothenburg, Sweden\\
              \email{mamiko@chalmers.se}
        \and
            Department of Space, Earth and Environment, Onsala Space Observatory, Chalmers University of Technology, SE-439 92 Onsala, Sweden % Sabine
         \and
             Institute of Astronomy, The University of Tokyo, 2-21-1 Osawa, Mitaka, Tokyo 181-0015, Japan % Kohno
             %\email{e}
             %\thanks{Here comse the thanks comment.}
        \and
            Leiden Observatory, Leiden University, PO Box 9513, 23000 RA Leiden, The Netherlands % Viti, van der Werf
        \and
             Transdisciplinary Research Area (TRA) ‘Matter’/Argelander-Institut für Astronomie, University of Bonn, Bonn, Germany % Viti
        \and
            Physics and Astronomy, University College London, London, UK % Viti
        \and
            Center for Interdisciplinary Exploration and Research in Astrophysics (CIERA) Northwestern University, Evanston, IL 60208, USA % Mark
        \and
            Observatoire de Paris, LERMA, Coll\'{e}ge de France, CNRS, PSL University, Sorbonne University, 75014 Paris, France % Combes
        \and
            Observatorio Astron\'{o}mico Nacional (OAN-IGN)- Observatorio de Madrid, Alfonso XII, 3, 28014-Madrid, Spain %Garcia-Burillo
         \and
            National Astronomical Observatory of Japan, 2-21-1 Osawa, Mitaka, Tokyo 181-8588, Japan % Harada
        \and
            William H. Miller III Department of Physics and Astronomy, Johns Hopkins University, Baltimore, MD 21218, USA % Justin
        \and
            Department of Astronomy, University of Wisconsin-Madison, 475 N Charter Street, Madison, WI 53706, USA % Jay
        \and
            Department of Astronomy, University of Virginia, 530 McCormick Road, Charlottesville, VA 22903, USA % Aaron
        \and
            National Radio Astronomy Observatory, 520 Edgemont Road, Charlottesville, VA, 22903, USA % Aaron
        \and
            Section of Astrophysics, Astronomy, and Mechanics, Department of Physics, National and Kapodistrian University of Athens, Panepistimioupolis Zografou, 15784 Athens, Greece % Kalliopi
        \and
            Finnish Centre for Astronomy with ESO (FINCA), University of Turku, Väisäläntie 20, FI-21500 Piikkiö, Finland % Jari
             }

   \date{Submitted May 20, 2025}

% \abstract{}{}{}{}{}
% 5 {} token are mandatory
 
  \abstract
   {Molecular gas plays a critical role in regulating star formation and nuclear activity in galaxies. Sulphur-bearing molecules, such as H$_2$S, are sensitive to the physical and chemical environments in which they reside and are potential tracers of shocked, dense gas in galactic outflows and active galactic nuclei (AGN).}
   {We aim to investigate the origin of H$_2$S emission and its relation to dense gas and outflow activity in the central regions of nearby infrared-luminous galaxies.} 
   {We present ALMA Band 5 observations of the ortho-H$_2$S 1$_{1,0}$ – $1_{0,1}$ transition in three nearby galaxies: NGC\,1377, NGC\,4418, and NGC\,1266. We perform radiative transfer modelling using RADEX to constrain the physical conditions of the H$_2$S-emitting gas and compare the results to ancillary CO and continuum data.}
   {We detect compact H$_2$S emission in all three galaxies, arising from regions smaller than $\sim$150 pc. The H$_2$S spectral profiles exhibit broad line wings, suggesting an association with outflowing or shocked gas. In NGC4418 H$_2$S also appears to be tracing gas that is counterrotating. A peculiar red-shifted emission feature may be inflowing gas, or possibly a slanted outflow. RADEX modelling indicates that the H$_2$S-emitting gas has high densities ($n_{\rm H_2} \gtrsim 10^7$ cm$^{-3}$) and moderately warm temperatures (40--200 K). The derived densities exceed those inferred from CO observations, implying that H$_2$S traces denser regions of the ISM.}
  % {The compact morphology and broad velocity components of the H$_2$S emission suggest that shocks associated with molecular outflows are the likely mechanism for releasing H$_2$S into the gas phase via sputtering or thermal desorption.}

	
	\maketitle

%%%%%%%%%%%%%%%%%%%%%%%%%%%%%%%%%%%%%%%%%%%%%%%%%%%%%%%%%%%%%%
\section{Introduction}
\label{sec:intro}
%-----------begin Table "Basic information of sources"------------------------------------------------
\begin{table*}[!h]
	\caption{Basic information of sources}             % title of Table
	\label{tab:Basic}      % is used to refer this table in the text
	\centering                          % used for centering table
	\begin{tabular}{l c c c c c c c c c}        % centered columns (4 columns)
		\hline\hline                 % inserts double horizontal lines
Object & R.A. & Dec. & D$_{\rm L}$ \tablefootmark{(a)} & L$_{\rm IR}$ \tablefootmark{(b)} & type & outflow \tablefootmark{(c)}\\
 & (J2000) & (J2000) & (Mpc) & (L$_{\rm \odot}$) &  &  \\
 \hline
NGC\,1377 & 03:36:39.1 & $-$20:54:08 & 21 & 1.3 $\times$ 10$^{10}$ & AGN? with CON & (1) \\
NGC\,4418 & 12:26:54.6 & $-$00:52:39 & 31.9 & 1.6 $\times$ 10$^{11}$ & AGN with CON & (2) \\
NGC\,1266 & 03:16:00.75 & $-$02:25:38.5 & 29.9 & 2.9 $\times$ 10$^{10}$  & AGN & (3)\\

		\hline
		\hline                                   
	\end{tabular}
	\tablefoot{
		\tablefoottext{a}{Distances taken from \citet{Sanders2003}.}
		\tablefoottext{b}{Infrared luminosities taken from  \cite{Sanders2003};}
		\tablefoottext{c}{References: (1) e.g. \cite{Aalto2012DetectionWind,Aalto2016},(2) e.g. \cite{Ohyama2019}, (3) e.g. \cite{Alatalo2011, Davis2012GeminiNGC1266}}}
            
\end{table*}

% Molecular gas plays a crucial role in triggering bursts of star formation and fueling supermassive black holes (SMBHs), particularly during collisions of gas-rich galaxies. These interactions can channel large quantities of material into the nuclei of luminous and ultraluminous infrared galaxies (LIRGs/ULIRGs) \citep{Sanders1996,Iono2009}. Molecular gas provides key insights into physical processes such as massive outflows driven by nuclear activity \citep[e.g.][]{Sturm2011, Feruglio2011, Chung2011, Aalto2012, Aalto2012b, Cicone2014, Sakamoto2014, Aalto2015a, Fluetsch2019, Veilleux2020CoolImplications}. These outflows regulate the growth of galactic nuclei, and the momentum they carry suggests that the central regions of these galaxies might be cleared of gas within just a few million years \citep[e.g.][]{Feruglio2011, Cicone2014}. The morphology of such molecular feedback varies widely, from wide-angle winds \citep{Veilleux2013} and gas entrained by radio jets \citep{Morganti2015, Garcia-Burillo2015High-resolution1614, Dasyra2016}, to tightly collimated molecular outflows \citep{Aalto2016, Sakamoto2017a, Barcos-Munoz2018, Falstad2019}.

Molecular gas plays a crucial role in star formation and fueling supermassive black holes (SMBHs), and bursts of star formation are particularly prevalent during collisions of gas-rich galaxies. Galaxy interactions can channel large quantities of material into the nuclei of luminous and ultraluminous infrared galaxies (LIRGs/ULIRGs) \citep{Sanders1996,Iono2009}. Furthermore, molecular gas provides key insights into physical processes such as massive outflows driven by nuclear activity \citep[e.g.][]{Sturm2011, Feruglio2011, Chung2011, Aalto2012, Aalto2012b, Cicone2014, Sakamoto2014, Aalto2015a, Fluetsch2019, Veilleux2020CoolImplications}. Such outflows regulate the growth of galactic nuclei, and the momentum they carry suggests that the central regions of these galaxies might be cleared of gas within just a few million years \citep[e.g.][]{Feruglio2011, Cicone2014}. But the morphology of such molecular feedback varies widely, from wide-angle winds \citep{Veilleux2013} and gas entrained by radio jets \citep{Morganti2015, Garcia-Burillo2015High-resolution1614, Dasyra2016}, to tightly collimated molecular outflows \citep{Aalto2016, Sakamoto2017a, Barcos-Munoz2018, Falstad2019}.

The ultimate fate of the molecular gas, however, remains unclear. One outstanding question is whether this gas escapes the galaxy or returns to fuel new episodes of activity \citep[e.g.][]{Pereira-Santaella2018, Lutz2020}. 
The nature of the molecular phase in outflows is also uncertain: do they consist of molecular clouds swept out of the circumnuclear disk or can molecular clouds form in situ within the outflow \citep[e.g.][]{Ferrara2016}?
%It is also uncertain whether the molecular clouds are simply swept up from a circumnuclear disk or formed in situ within the outflow \citep[e.g.][]{Ferrara2016}. 
Moreover, it is not known whether clouds in the outflow expand and evaporate or condense and potentially form stars while embedded in the outflow \citep{Maiolino2017}. To adress these questions, we need detailed studies of the physical conditions of the molecular gas in outflows to understand its origins, evolution, and how these processes are linked to the driving mechanisms of the outflows.

Besides CO, a range of molecular species, including HCN, HCO$^+$, HNC, CN, and HC$_3$N, are essential tools for probing cloud properties in external galaxies. Due to their high dipole moments, these molecules typically require high gas densities ($n > 10^4$ cm$^{-3}$) to produce emission from even their low-J rotational transitions. Consequently, they are more directly associated with dense star-forming regions than, for example, the lower-$J$ transitions of CO, as shown by \citet{Gao2004}.

The detection of luminous HCN and CN emission within molecular outflows, %often associated with active galactic nuclei (AGN), 
indicates the presence of significant amounts of dense molecular gas \citep[e.g.][]{Aalto2012, Sakamoto2014, Matsushita2015, Garcia-Burillo2015High-resolution1614, Privon2015, Walter2017,Harada2018, Cicone2020, Saito2022AGN-drivenMolecules}. However, it should also be noted that faint but widespread HCN emission may arise from more diffuse molecular gas as well \citep[e.g.][]{Nishimura2017}.

To discern the origin of this dense gas, we need tracers sensitive to specific physical and chemical environments. Being one of the most chemically reactive elements \citep[e.g.][]{Mifsud2021SulfurStudies}, sulphur exhibits strong sensitivity to the kinetic and thermal state of the gas. Sulphur-bearing molecules, such as H$_2$S, SO, and SO$_2$, tend to be unusually abundant in regions with high-mass star formation (around 0.1 pc size region), particularly where shocks or elevated temperatures are present \citep[e.g.][]{Mitchell1984, MInh1990}. Observations of these sulphur-bearing species can therefore help constrain the properties of dense gas in outflows.

H$_2$S is thought to be the dominant sulphur-bearing molecule on icy grain mantles, as sulphur readily undergoes hydrogenation on dust surfaces, despite its lack of detection in interstellar ices \citep{Charnley1997, Wakelam2004, Viti2004, Bariosco2024TheIces}. Consequently, in quiescent star-forming regions, H$_2$S is generally found in low gas-phase abundances due to its depletion onto grains.
However, shocks can liberate H$_2$S from the grains, injecting it into the gas phase. 
This is supported by theoretical studies \citep[e.g.][]{Woods2015}, showing clear evidence for enhanced H$_2$S abundances in warm and shocked gas. 
In particular, \citet{Holdship2017} showed that the propagation of a 
%continuous (C-type, need definition also for J-shock as comparison)
C-type shock through a cloud can efficiently form H$_2$S, resulting in a significant abundance increase in the post-shock gas. The chemical evolution of H$_2$S, along with its connection to grain processing, shares similarities with water (H$_2$O) chemistry. However, an observational advantage is that H$_2$S has a ground-state transition at $\lambda = 2$ mm, which is accessible to ground-based telescopes.

For the reasons mentioned above, H$_2$S emerges as a sensitive tracer for studying outflows and obscured nuclear activity in galaxies.\citep{Sato2022APEXOutflows}. 
Observations of the ground-state transition of H$2$S (1$_{1,0}$-1$_{0,1}$ at 168.7 GHz) toward 12 nearby luminous infrared galaxies (LIRGs) detected emission in 9 sources, with simultaneous HCN and HCO$^+$ line measurements. While H$_2$S abundance ratios relative to HCN and HCO$^+$ showed no clear connection to galactic-scale molecular outflows, its line luminosity correlated more strongly with outflow mass than traditional CO tracers. This suggests H$_2$S enhancement likely arises from small-scale shocks in nuclear regions rather than large-scale outflow activity. To resolve the emission sources, follow-up Atacama Large Millimeter Array (ALMA) observations achieved 0".1-1".2 spatial resolution toward three galaxies, enabling direct comparison between H$_2$S distribution and outflows.
%We observed the ground-state transition of H$_2$S toward a sample of 12 nearby LIRGs, using a beam size of approximately 37" at 2~mm wavelength. The observations were configured to simultaneously detect HCN and HCO$^+$ emission lines. We detected H$_2$S emission in 9 out of the 12 galaxies. By defining the enhancement of H$_2$S abundance through the line ratios relative to HCN and HCO$^+$, we found no clear correlation between these ratios and the reported presence or absence of galactic-scale molecular outflows. However, the H$_2$S line luminosity appeared to correlate more tightly with the molecular outflow mass than CO does. We concluded that H$_2$S emission is likely enhanced by small-scale shocks within the nuclear regions, suggesting that higher spatial resolution observations are required to pinpoint the exact location of the H$_2$S-emitting gas. Here we present the results of Atacama Large Millimeter Array (ALMA) observations of the 168.7 GHz 1$_{1,0}$-1$_{0,1}$ transition of H$_2$S with spatial resolution of 0".1-1".2 towards a sample of three galaxies to study the relationship between H$_2$S emission and the outflow activity of the galaxies. 

This paper is organized as follows: Section \ref{sec:obs} describes the galaxy sample and ALMA observations. Section \ref{sec:results} presents the observational results. In Section \ref{sec:discu}, we discuss the possible mechanisms for the formation of Gas-phase H$_2$S and origin of H$_2$S in the observed galaxies.. Finally, Section \ref{sec:conc} summarises our conclusions and provides future directions for understanding the role of H$_2$S as a tracer of dense gas and/or small scale shocks.
%%%%%%%%%%%%%%%%%%%%%%%%%%%%%%%%%%%%%%%%%%%%%%%%%%%%%%%%%%%%%%

%%%%%%%%%%%%%%%%%%%%%%%%%%%%%%%%%%%%%%%%%%%%%%%%%%%%%%%%%%%%%%
\section{Observations and data reduction}
\label{sec:obs}

\subsection{Sample selection}
\label{subsec:sample}

\citet{Sato2022APEXOutflows} reported the detection of H$_2$S emission in nine infrared-luminous galaxies. These galaxies encompass a wide variety of evolutionary stages and environmental conditions. To further explore the evolution of galaxies with dusty dense cores, we selected a subgroup of three nearby galaxies, each exhibiting molecular outflows but at different stages of interaction: NGC\,1377, NGC\,4418, and NGC\,1266. The target properties are given in Table \ref{tab:Basic}.
All of them are relatively infrared luminous ($L_{\rm IR} =  10^{10} - 10^{11} L_{\odot}$), suggesting that they host a dense dusty core. 
% These galaxies also vary in central dust content. 
In the following, we provide a brief overview of the key characteristics of these three galaxies.

%: a merging galaxy (N1377), an interacting galaxy (N4418), and an isolated galaxy (N1266). 

%- N1377 collimated molecular outflow, N(H2)10e23, Lir=10e10, obscured nucleus
NGC\,1377 (z=0.00598) is a nearby lenticular galaxy at an estimated distance of 21 Mpc (1" = 102 pc) and has an IR luminosity of $L_{\rm IR} = 1.3 \times 10^{10} L_{\odot}$ \citep{Sanders2003}. 
A powerful molecular jet (with a mass outflow rate of 8 - 35 $\rm M_{\odot} yr^{-1}$) was found \citep[e.g.][]{Aalto2012DetectionWind,Aalto2016,Aalto2020ALMA1377}. It is also in the list of galaxies that host a Compact Obscured Nucleus (CON, \citet{Falstad2021}), which is identified with extremely high nuclear gas column density (N$_{H_2} > 10^{24}$ cm$^{-2}$).
The nucleus is above the radio-FIR correlation \citep{Spoon2006MID-IRWIDTH}, though it is unclear whether the nucleus hosts a nascent starburst \citep{Roussel2003NASCENT1,Roussel2006THEOBSERVATIONS} or a radio-quiet AGN due to the extreme dust obscuration in the centre \citep{Imanishi2006INFRARED1}.
%A nascent starburst (Roussel et al. 2003,2006) or a radio-quiet AGN \citep{Imanishi2006INFRARED1} are proposed be at the central region, due to its deviation from the radio-FIR correlation and the dust-enshrouded central region \citep{Spoon2006MID-IRWIDTH}.

%- N4418 CON, molecular outflow, N(H2) >10e24, LIRG (Lir=10e11)
NGC\,4418 is one of the nearest (z=0.00727, 1" = 150 pc) CON-hosting galaxies which is also a LIRG ($L_{\rm IR} = 1.6 \times 10^{11} L_{\odot}$ \citep{Sanders2003}). It is composed of an interacting system with the nearby galaxy VV~655.
A kpc-scale polar outflow has been reported \citep[e.g.][]{Sakamoto2013,Ohyama2019}.

%- N1266 massive AGN-driven molecular outflow, non-interacting galaxy to host a molecular outflow. Lir=10e10
NGC\,1266 (z=0.00724, 1" = 145 pc) is a non-interacting nearby S0 galaxy with a dense and compact molecular nucleus and a massive AGN-driven molecular outflow from its central region \citep{Alatalo2011,Alatalo2015Escape231}.
Similar to NGC\,4418, the nature of its nuclear activity remains ambiguous, with both AGN and compact starburst scenarios being viable explanations for its energetic emission and outflow properties \citep{Nyland2013Detection1266,Alatalo2015Escape231}. 
Furthermore, shocks are widespread throughout the galaxy's interstellar medium, evidenced by highly excited H$_2$ emission and disturbed ionized gas kinematics, likely driven by interactions between the AGN-driven outflow and dense circumnuclear gas \citep{Pellegrini2013ShockGalaxy,Otter2024PullingGemini-NIFS}. 
% * similar to 4418, it’s unclear if the center is an AGN or starburst (see Nyland13, Alatalo15)
% * there are shocks throughout the galaxy, see Pellegrini13 and Otter24

\subsection{ALMA Observations}
\label{subsec:obs}
% - H$_2$S 168 obs by PI:Sato data.
The observations in the ortho-H$_2$S 1$_{1,0}$-1$_{0,1}$ line were carried out with ALMA in October 2018 for NGC\,1377 and NGC\,1266  (2018.1.00423.S) and
% - H$_2$S 168 obs by PI:Stanley data.
in August 2019 for NGC\,4418 (2018.1.00939.S), respectively,
% - Obs settings/configurations for H$_2$S168 data. Receiver info. Reference to band5 receiver.
using the Band 5 receivers \citep{Belitsky2018SEPIATelescope}. 
The correlator was set up with two 1.875~GHz-wide (480~channels) spectral windows for NGC\,1377 and NGC\,1266, four 1.875~GHz-wide (240~channels) spectral windows for NGC4418, in a way that one of the windows centred at the redshifted H$_2$S 1$_{1,0}$-1$_{0,1}$ line ($\nu_{\rm rest}$ = 168.763~GHz, $E_{\rm u}$ = 27.9~K).

% - H$_2$S 216 archival data in the same style as above, for each source.
Complementary data of the para-H$_2$S 2$_{2,0}$-2$_{1,1}$ line ($\nu_{\rm rest}$ = 216.710~GHz, $E_{\rm u}$ = 84.0~K) were obtained from the ALMA archive : project code 2018.1.01488.S for NGC\,1377, 2012.1.00375.S for NGC\,4418, and 2011.0.00511.S for NGC\,1266 in band 6 \citep{Ediss2004ALMAPerformance}.

% - CO65 data for N1377, project id, obs configurations mm.
Finally, data of the CO 6-5 transition were included in the analysis.
The ALMA observations of NGC\,1377 in the CO (6-5) line were carried out in December 2018 to January 2021 for project code 2018.1.01488.S.
The Band 9 receiver was used \citep{Baryshev2015TheLight}. The redshifted CO (6-5) line was covered in one of the four 1.875~GHz-wide (960 channels) spectral windows, centred at 687.491~GHz.

% - Data reduction explanation.
\subsection{Data reduction}
\label{subsec:reduction}
For each source, we used the calibrated data sets provided by the ALMA calibration pipeline. We fit and subtracted the continuum in the $uv$ plane through the CASA task "uvcontsub". We used a zeroth-order polynomial for the fit, and estimated the continuum emission in the line-free frequencies.
%in the following ranges: 165.0 > $\nu_{\rm obs} [\rm GHz]$ > 168.7 for NGC\,1377, 164.8 > $\nu_{\rm obs} [\rm GHz]$ > 168.5 for NGC\,1266, and 167.1 > $\nu_{\rm obs} [\rm GHz]$ > 170.7 and 179.3 > $\nu_{\rm obs} [\rm GHz]$ > 182.9 for NGC\,4418.
Images of the sources were produced using the "tclean" task in CASA with natural weighting. A spectral binning of $\Delta v$=20 km s$^{-1}$ was applied. The typical angular resolution is 0."2 - 0."5.
%(until here)
% - lead to table 1.
A summary of the observations is given in Table \ref{tab:obs}.

%-----------begin Table "Description of the ALMA observations and Cont results"----------
\begin{table*}[!h]
	\caption{Description of the ALMA observations and the continuum results}             % title of Table
	\label{tab:obs}      % is used to refer this table in the text
	\centering                          % used for centering table
	\begin{tabular}{l c c c c c c c c c c c}        % centered columns (4 columns)
		\hline\hline                 % inserts double horizontal lines $L_{\mathrm{IR}}$
Object & Project ID & $\nu_{\rm obs}$ \tablefootmark{(a)}  & $b_{\mathrm{min}}$, $b_{\mathrm{max}}$ & $t_{\mathrm{int}}$ \tablefootmark{(b)}& $\theta_{maj} \times \theta_{min}$ & $I_{\mathrm{cont,peak}}$ &  $rms_{\mathrm{cont}}$ \tablefootmark{(c)} & $S_{\mathrm{cont}}$  \tablefootmark{(d)} \\
 &  & [GHz] & [m] & [hrs] &  [arcsec$^2$] & [mJy~beam$^{-1}$] & [mJy~beam$^{-1}$] & [mJy] \\
 \hline
NGC\,1377 & 2018.1.00423.S & 167.76 & 67/585 & 1.1  & 0.$^"$51 $\times$ 0.$^"$43 & 0.198 & 0.011 & 0.30$\pm$0.013 \\
         & 2018.1.01488.S & 216.71 & 15/13900 & 1.1  & 0.$^"$10 $\times$ 0.$^"$08 & 0.178 & 0.018 & 0.28$\pm$0.02 \\
         & 2018.1.01488.S & 691.47 & 37/1154 & 1.1  & 0.$^"$19 $\times$ 0.$^"$15 & 9.65 & 0.407 & 31.5$\pm$2.4 \\
NGC\,4418 & 2018.1.00939.S & 167.575 & 155/1500 & 0.4 & 0.$^"$21 $\times$ 0.$^"$16 & 18.4 & 0.076 & 23.4$\pm$0.15 \\
         & 2012.1.00377.S & 216.71 & 15/1600 & 0.1 & 0.$^"$37 $\times$ 0.$^"$23 & 24.1 & 0.643 & 45.0$\pm$1.18\\
NGC\,1266 & 2018.1.00423.S & 167.55 & 66/565 & 0.9 &0.$^"$52 $\times$ 0.$^"$47 & 2.59 & 0.013 & 3.84$\pm$0.11 \\
         & 2011.1.00511.S & 216.71 & 21/384 & 1.4  & 1.$^"$26 $\times$ 0.$^"$82 & 5.22 & 0.120 & 5.74$\pm$0.24 \\
\hline\hline
	\end{tabular}
	\tablefoot{	
        \tablefoottext{a}{Central frequency of the spectral window which include the targeted line.};
        \tablefoottext{b}{On source observation time};
		\tablefoottext{c}{ Noise level determined using CLASS package in GILDAS (http://www.iram.fr/IRAMFR/GILDAS).};
        \tablefoottext{d}{Flux density determined with CASA from the region with over 3 $\sigma$ emission for each source.}
	}
\end{table*}
%%%%%%%%%%%%%%%%%%%%%%%%%%%%%%%%%%%%%%%%%%%%%%%%%%%%%%%%%%%%%%

%%%%%%%%%%%%%%%%%%%%%%%%%%%%%%%%%%%%%%%%%%%%%%%%%%%%%%%%%%%%%%
\section{Results}
\label{sec:results}
% Overview of the shown results
The H$_2$S 1$_{1,0}$-1$_{0,1}$ transition line is detected in emission toward all three galaxies.
Spectra and integrated intensity maps are shown in Figures \ref{fig:1377H2S11}, \ref{fig:4418H2S11}, and \ref{fig:1266H2S11}, respectively for each galaxy.
We also present the detection of the H$_2$S 2$_{2,0}$-2$_{1,1}$ transition line in Figs. \ref{fig:1377H2S22}, \ref{fig:4418H2S22}, and \ref{fig:1266H2S22}, for which we obtained the data from the ALMA archive.
Toward NGC\,1377, archival data also show a detection of CO 6-5. The spectra, the continuum emission, and the integrated intensity map for this added data set are shown in Fig. \ref{fig:1377CO} in the same manner as for H$_2$S.\\
% Explanation of the spectra
Panels (c) in Figs. \ref{fig:1377H2S11} - \ref{fig:1266H2S22} show the spectra taken from the region within the 3 $\sigma$ contour in the moment 0 image.
Panels (d) in Figs. \ref{fig:1377H2S11} - \ref{fig:1266H2S22}  show the spectra taken from the central pixel.
The H$_2$S $1_{1,0} - 1_{0,1}$ and H$_2$S $2_{2,0} - 2_{1,1}$ lines with higher signal-to-noise ratios show profiles that appear to include high-velocity wings.% is this true for all the sources? or only one or two of them?
We found that a single Gaussian component is insufficient to adequately fit these lines, but another component is required.
The individual Gaussian fits (dashed and dotted lines), along with the combined fits (solid grey line), are shown with the spectra of the galaxies (black histogram) in all spectra except the ones for H\(_2\)S $2_{2,0} - 2_{1,1}$ in NGC\,1377 and NGC\,1266, which lines have a lower signal-to-noise ratio and could be adequately fitted with only one Gaussian component. The red dotted line indicates the 1 $\sigma$ noise level.

In Table \ref{tab:res2}, the parameters of the Gaussian fits are listed together with the velocity-integrated flux densities.\\% by CASA.\\
% Explanation of the maps
Velocity-integrated intensity maps were created by integrating the emission over different velocity ranges, and these are shown as contours of different colours in the images. The integration velocity ranges for each source are further discussed below.\\

\subsection{Line emission for individual sources}
\label{subsec:line}
\subsubsection{NGC\,1377}
\label{subsubsec:1377}

% Line profiles
%This is the only galaxy of the three presented in this work for which CO 6-5 emission is available. 
The line profiles of H$_2$S 1$_{1,0}$-1$_{0,1}$ exhibit clear high-velocity wings that cannot be fitted with a single Gaussian component. A double Gaussian fit shows that one of the Gaussians is narrow and brighter, which we refer to as the core component, and the other is broader and less bright, which we refer to as the wings (see Table \ref{tab:res2}).
From the Gaussian fits to H$_2$S 1$_{1,0}$-1$_{0,1}$ it is found that the core component exhibits a line width of $\sim$ 60~km~s$^{-1}$, while the broader component has a width of $\sim$ 130 km s$^{-1}$.

Forcing the central velocity to be the same as for the H$_2$S 1$_{1,0}$-1$_{0,1}$ line, the H$_2$S 2$_{2,0}$-2$_{1,1}$ line exhibits a line width of $\sim$ 70 km s$^{-1}$. This is the width for the whole line because the signal-to-noise ratio is too low to meaningfully separate it into narrow and broad components.
The CO 6-5 line also needs a double Gaussian fit for its line profile, exhibiting a line width of $\sim$ 95~km~s$^{-1}$ for the line core component and $\sim$ 175~km~s$^{-1}$ for the wing component.

% about the morphology
We find a centrally peaked structure in the velocity integrated maps for all three lines (Figs.\ref{fig:1377H2S11} - \ref{fig:1377CO}). 
In H$_2$S 1$_{1,0}$-1$_{0,1}$, the line core emission (corresponding to the narrow Gaussian component) comes from a region of 1$^{\prime\prime}$.4 (around 150pc) diameter from the centre, with a slight elongation in the west-east direction.
The location of the line wing components are overlapping on top of the continuum peak of the galaxy, which is more compact in size than the emission corresponding to the line core component. The blue-shifted emission is extended toward the north-east of the galaxy centre,
while the red-shifted emission is confined at the central beam-size region.
The H$_2$S 2$_{2,0}$-2$_{1,1}$ emission comes from a very compact region (0$\rlap{.}^{\prime\prime}$17 $\sim$ 17 pc). \\
%Note that the black contour shows the integrated intensity from the whole line. \\
In CO 6-5, the line core emission comes from a slightly more extended region (1$^{\prime\prime}$.6 diameter) than H$_2$S 1$_{1,0}$-1$_{0,1}$ but has similar tendency to extend more in the east-west direction compared to north-south. The blue- and red-shifted emissions are extended toward the opposite direction to each other, north-east and south-west, respectively. Both are within the range of the core component. 
Fig. \ref{fig:1377CO32} shows a comparison between CO 6-5 and CO 3-2 high velocity gas in NGC\,1377. It shows an excitation gradient along the axis of the outflow with CO 6-5 tracing warmer and/or dense gas at the beginning of the outflow (30-50~pc). It also shows a slight deviation from the indicated jet axis with a more north-south orientation. This is consistent with the notion of a precessing or swirling jet as discussed in \citet{Aalto2016, Aalto2020ALMA1377}.

\subsubsection{NGC\,4418}
\label{subsubsec:4418}
\begin{figure*}
    \centering
    \includegraphics[width=0.8\textwidth]{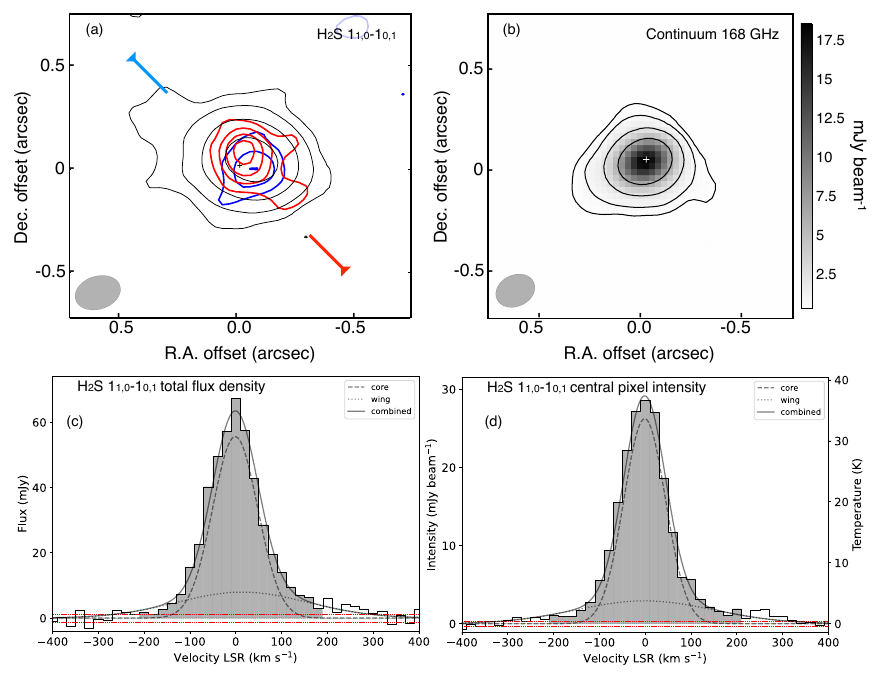}
    \caption{The results of 168~GHz observations towards NGC\,4418. (a) The velocity-integrated line emission maps of H$_2$S 1$_{1,0}$-1$_{0,1}$: black contours: -100 km s$^{-1}$ to 100 km s$^{-1}$ (3,9,27,81, and 243 $\sigma$ where $\sigma_{\rm core} = 2.63 \times 10^{-2}$ Jy~beam$^{-1}$ km s$^{-1}$), blue: -200 km s$^{-1}$ to -100 km s$^{-1}$, red: 100 km s$^{-1}$ to 200 km s$^{-1}$ (3,4,6, and 9 $\sigma$ where $\sigma_{\rm red} = 2.41 \times 10^{-2}$ Jy~beam$^{-1}$ km s$^{-1}$ and $\sigma_{\rm blue} = 2.28 \times 10^{-2}$ Jy~beam$^{-1}$ km s$^{-1}$, respectively).  The beam size is shown at the bottom-left in grey-filled circle. The blue and red line segments indicate the position angle (PA) of the disk \citep[$\sim 45^{\circ}$, ][]{Costagliola2013}.
    (b) The Continuum emission at 168~GHz in greyscale and contours: (3,9,27, and 81 $\sigma$ where $\sigma_{\rm cont} = 7.6 \times 10^{-5}$ Jy~beam$^{-1}$).
    (c) Spectrum in the unit of the flux density within the emitting region (over 3 $\sigma$ in moment 0 image) against the line velocity offset at the systemic velocity. The black histogram shows the observation result. The red-dotted line indicates the 1 $\sigma$ rms level. The black dashed line:narrow line component, dotted line: broad line component, and solid line: two components combined.
    (d) Spectrum in unit of the mean brightness at the peak intensity pixel against the line velocity.}
    \label{fig:4418H2S11}
\end{figure*}

% Line profiles
Similar to NGC\,1377, the line profiles of H$_2$S 1$_{1,0}$-1$_{0,1}$ in NGC\,4418 also exhibit clear high-velocity wings that cannot be fitted with a single Gaussian component. 
From the two Gaussian fits to H$_2$S 1$_{1,0}$-1$_{0,1}$ it is found that the core component has a line width of $\sim$ 107~km~s$^{-1}$, while the broader component has a width of $\sim$ 353 km s$^{-1}$ (Fig.\ref{fig:4418H2S11}(d)).

The line profile of the H$_2$S 2$_{2,0}$-2$_{1,1}$ line in NGC\,4418 is asymmetric and we suspect that the excess in the blue-shifted emission is a blending from an emission line of another species. Indeed NGC\,4418 has a very rich spectrum \citep[e.g.][]{Costagliola2015}. One of the potential candidate for this blended feature is the $^{\rm 13}$CN (N= 2- 1, J=3/2-3/2, F1= 1- 2, F= 0- 1) line. Assuming that, we tried to fit three Gaussian components: two with the central velocity fixed to what we derived for the H$_2$S 1$_{1,0}$-1$_{0,1}$ transition line (core and wings components), and one with the central velocity around 200~km~s$^{-1}$ offset from the central velocity of the H$_2$S line towards lower velocities. 
Thus we found that the core component has a similar line width as for H$_2$S 1$_{1,0}$-1$_{0,1}$ line, $\sim$ 109~km~s$^{-1}$, and the broader component has a width of $\sim$ 236 km s$^{-1}$.

\begin{figure*}
    \centering
    \includegraphics[width=0.8\textwidth]{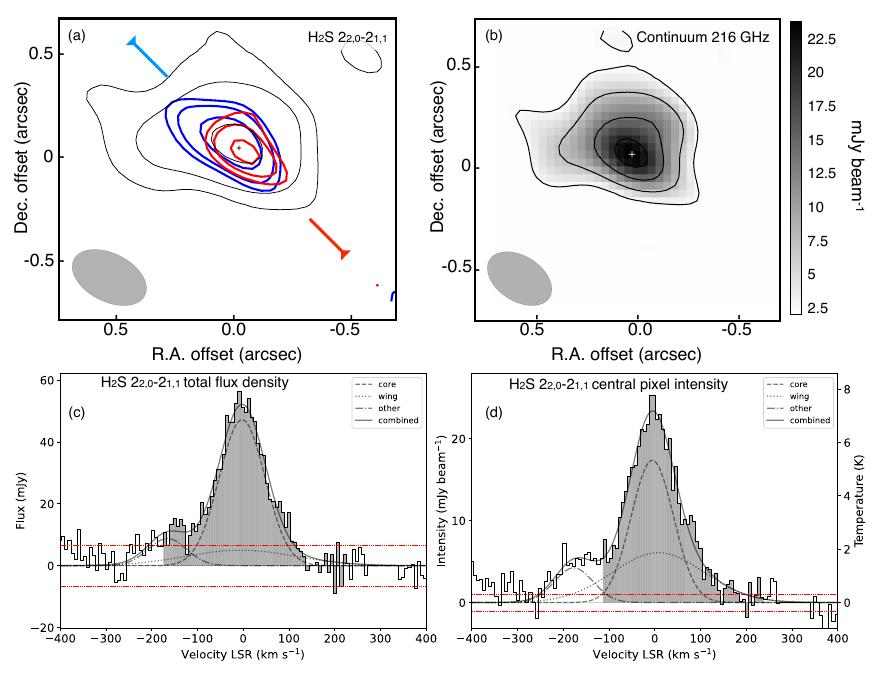}
    \caption{The results of 216~GHz observations towards NGC\,4418. (a) The velocity-integrated line emission maps of 2$_{2,0}$-2$_{1,1}$: black contours: -100 km s$^{-1}$ to 100 km s$^{-1}$ (3,9, and 27 $\sigma$ where $\sigma_{\rm core} = 6.7 \times 10^{-2}$ Jy~beam$^{-1}$ km s$^{-1}$), blue: -120 km s$^{-1}$ to -100 km s$^{-1}$, red: 100 km s$^{-1}$ to 200 km s$^{-1}$ (3,4, and 6 $\sigma$ where $\sigma_{\rm red} = 4.2 \times 10^{-2}$ Jy~beam$^{-1}$ km s$^{-1}$ and $\sigma_{\rm blue} = 1.9 \times 10^{-2}$ Jy~beam$^{-1}$ km s$^{-1}$, respectively).  The lines and symbol conventions are the same as in Fig.\ref{fig:4418H2S11}.
    (b) The Continuum emission at 216~GHz in greyscale and contours: (3,9,27, and 81 $\sigma$ where $\sigma_{\rm cont} = 6.4 \times 10^{-4}$ Jy~beam$^{-1}$).
    (c) Spectrum in the unit of the flux density within the emitting region (over 3 $\sigma$ in moment 0 image) against the line velocity offset at the systemic velocity. The black histogram shows the observation result. The red-dotted line indicates the 1 $\sigma$ rms level. The black dashed line:narrow line component, dotted line: broad line component, dash-dotted line: other species (possibly 13CN line), and solid line: three components combined.
    (d) Spectrum in unit of the mean brightness at the peak intensity pixel against the line velocity.}
    \label{fig:4418H2S22}
\end{figure*}

% about the morphology
We find a centrally peaked structure in the velocity integrated maps for both lines (Fig.\ref{fig:4418H2S11} and \ref{fig:4418H2S22}). 
In H$_2$S 1$_{1,0}$-1$_{0,1}$, the line core emission comes from a region of 1$^{\prime\prime}$.0 (around 150pc) diameter from the centre.
There is a slight elongation in the north-east direction which is not detected in the continuum emission.
The location of the line wing components are overlapping on top of the continuum peak of the galaxy, which is more compact in size than the emission corresponding to the line core component. The blue-shifted emission is extended toward the south-east of the galaxy centre and the red shifted emission has a clear emission toward the south-west of the centre.

The H$_2$S 2$_{2,0}$-2$_{1,1}$ emission also comes from a similar region as H$_2$S 1$_{1,0}$-1$_{0,1}$ (1$^{\prime\prime}$). 
The north-east elongation is more apparent. The line wings components are coming from the central beam region. The whole line emission shows almost the same distribution as the continuum emission.
\\

\subsubsection{NGC\,1266}
\label{subsubsec:1266}
\begin{figure*}
    \centering
    \includegraphics[width=0.8\textwidth]{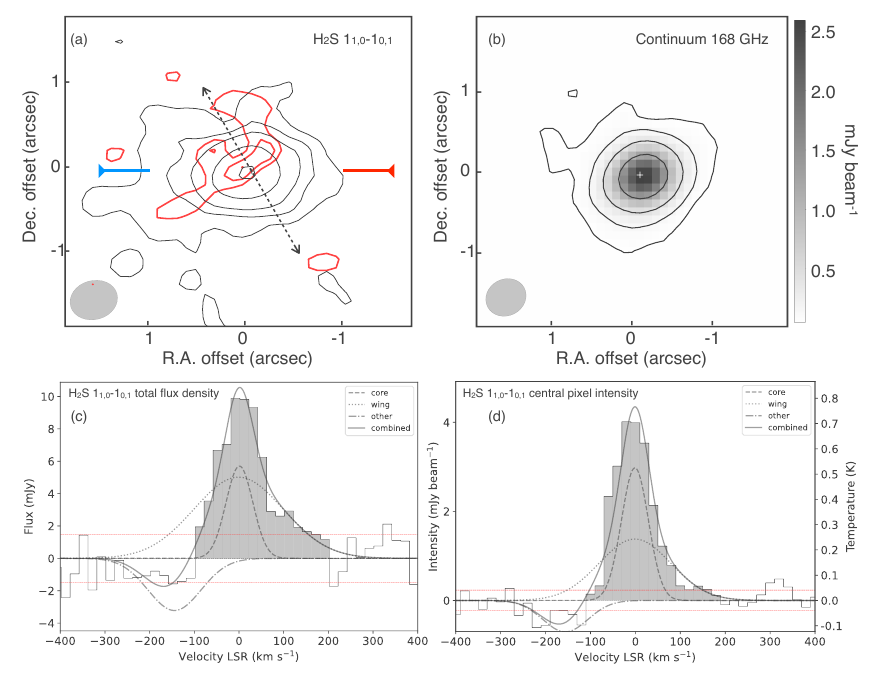}
    \caption{The results of 168~GHz observations towards NGC\,1266. (a) The velocity-integrated line emission maps of H$_2$S 1$_{1,0}$-1$_{0,1}$:black contours: -70 km s$^{-1}$ to 70 km s$^{-1}$ (3,6,9, and 18 $\sigma$ where $\sigma_{\rm core} = 1.08 \times 10^{-2}$ Jy~beam$^{-1}$ km s$^{-1}$), blue: -100 km s$^{-1}$ to -70 km s$^{-1}$, red: 70 km s$^{-1}$ to 200 km s$^{-1}$ (3,4, and 5 $\sigma$ where $\sigma_{\rm red} = 7.5 \times 10^{-3}$ Jy~beam$^{-1}$ km s$^{-1}$ and $\sigma_{\rm blue} = 5.7 \times 10^{-3}$ Jy~beam$^{-1}$ km s$^{-1}$, respectively).  The beam size is shown at the bottom-left in grey-filled circle. The blue and red line segments indicate the position angle (PA) of the disk ($\sim 90^{\circ}$) and direction of the molecular outflows detected in CO~2-1 (PA $\sim 30^{\circ}$) \citep{Alatalo2011}.
    (b) The Continuum emission at 168~GHz in greyscale and contours: (3,9,27, and 81 $\sigma$ where $\sigma_{\rm cont} = 1.3 \times 10^{-5}$ Jy~beam$^{-1}$).
    (c) Spectrum in the unit of the flux density within the emitting region (over 3 $\sigma$ in moment 0 image) against the line velocity offset at the systemic velocity. The black histogram shows the observation result. The red-dotted line indicates the 1 $\sigma$ rms level. The black dashed line:narrow line component, dotted line: broad line component, dash-dotted line: absorption component, and solid line: three components combined.
    (d) Spectrum in unit of the mean brightness at the peak intensity pixel against the line velocity.}    
    \label{fig:1266H2S11}
\end{figure*}

% Line profiles
The H$_2$S 1$_{1,0}$-1$_{0,1}$ line in NGC\,1266 exhibits a clear red-shifted emission. Such a line wing emission was not detected on the blue-shifted side of the line. 
There seems to be no candidate species that have a transition on the red shifted side of the H$_2$S 1$_{1,0}$-1$_{0,1}$ line (100-200 km s$^{-1}$).
So we assume that the red-shifted emission is part of the  H$_2$S 1$_{1,0}$-1$_{0,1}$ line.
It looks from the line profile (Fig.\ref{fig:1266H2S11}(c) and (d)) that the line wing component on the blue-shifted side are absorbed. Looking at the moment 0 map (Fig.\ref{fig:1266H2S11} (a)), the blue-shifted emission is not detected over 3$\sigma$. 
This type of line profile is consistent with an expanding gas observed against a warmer background continuum source. The continuum emission arises from a compact, point-like source (Fig.~\ref{fig:1266H2S11}(b)). If the absence of blue-shifted emission is due to absorption, this absorption should be stronger at the central pixel than in the spectrum integrated over the extended emission region. The intensity profile extracted from the central pixel shows a minimum of approximately 0.7mJybeam$^{-1}$, whereas the depth is only about 0.1mJybeam$^{-1}$ when averaged over the 3$\sigma$ emitting region (estimated by dividing the total flux in Fig.~\ref{fig:1266H2S11}(c) by the solid angle of the emission region). This difference is consistent with absorption occurring against a compact continuum source.
Thus we apply three Gaussian fittings, two with the same central velocity and one with the central velocity as a free parameter, assuming that the blue-shifted component of the line is absorbed due to the warm dust continuum at the centre. 
A line width of $\sim$ 71~km~s$^{-1}$ for the core component with a width of $\sim$ 188 km s$^{-1}$ for the wing component could potentially fit the line profile.

The H$_2$S 2$_{2,0}$-2$_{1,1}$ line in NGC\,1266 was detected at the edge of the observed frequency band and we can not see the full width of the line, unfortunately. Thus we fit one Gaussian and obtained an estimation for a line width of $\sim$ 50~km~s$^{-1}$.

\begin{figure*}
    \centering
    \includegraphics[width=0.8\textwidth]{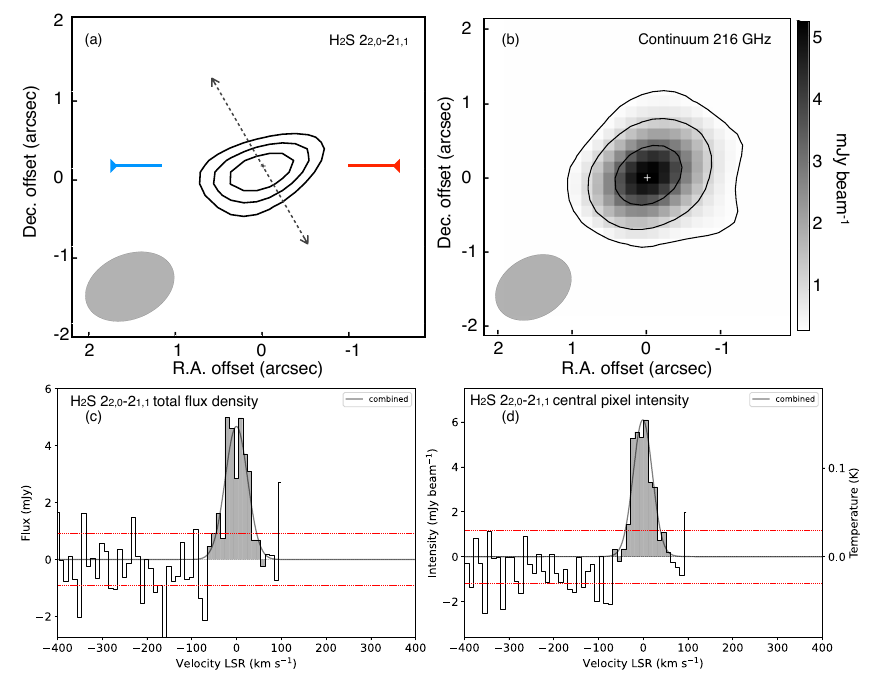}
    \caption{The results of 216~GHz observations towards NGC\,1266. (a) The velocity-integrated line emission maps of 2$_{2,0}$-2$_{1,1}$: black contours: -70 km s$^{-1}$ to 70 km s$^{-1}$ (3,6, and 9 $\sigma$ where $\sigma_{\rm core} = 1.08 \times 10^{-2}$ Jy~beam$^{-1}$ km s$^{-1}$). The lines and symbol conventions are the same as in Fig.\ref{fig:1266H2S11}.
    (b) The Continuum emission at 216~GHz in greyscale and contours: (3,9, and 27 $\sigma$ where $\sigma_{\rm cont} = 1.2 \times 10^{-4}$ Jy~beam$^{-1}$).
    (c) Spectrum in the unit of the flux density within the emitting region (over 3 $\sigma$ in moment 0 image) against the line velocity offset at the systemic velocity. The black histogram shows the observation result. The red-dotted line indicates the 1 $\sigma$ rms level. The black solid line: singls Gaussian fit.
    (d) Spectrum in unit of the mean brightness at the peak intensity pixel against the line velocity.}
    \label{fig:1266H2S22}
\end{figure*}

% about the morphology
We find a centrally peaked structure in the velocity integrated maps for both lines (Fig.\ref{fig:1266H2S11}(a) and Fig. \ref{fig:1266H2S22}(a)). 
In H$_2$S 1$_{1,0}$-1$_{0,1}$, the line core emission comes from a region of 2$^{\prime\prime}$.0 (around 300~pc) diameter from the centre.
Line core component is distributed in a slightly larger region than what the continuum emission is. 
The red-shifted component is detected from the nucleus to the north-east of the nucleus only in  H$_2$S 1$_{1,0}$-1$_{0,1}$. This direction agrees with the CO observations by \citet{Alatalo2011}.

The H$_2$S 2$_{2,0}$-2$_{1,1}$ emission is coming from a smaller region ($\sim$ 1$^{\prime\prime}$) than that of H$_2$S 1$_{1,0}$-1$_{0,1}$. The emitting region is basically the observed beam size (1$^{\prime\prime}$.26 $\times$ 0$^{\prime\prime}$.82), thus it is not resolved.
\\

%-----------begin Figures : Spectra & Maps for three sources----------------------------------

\begin{figure*}
    \centering
    \includegraphics[width=0.8\textwidth]{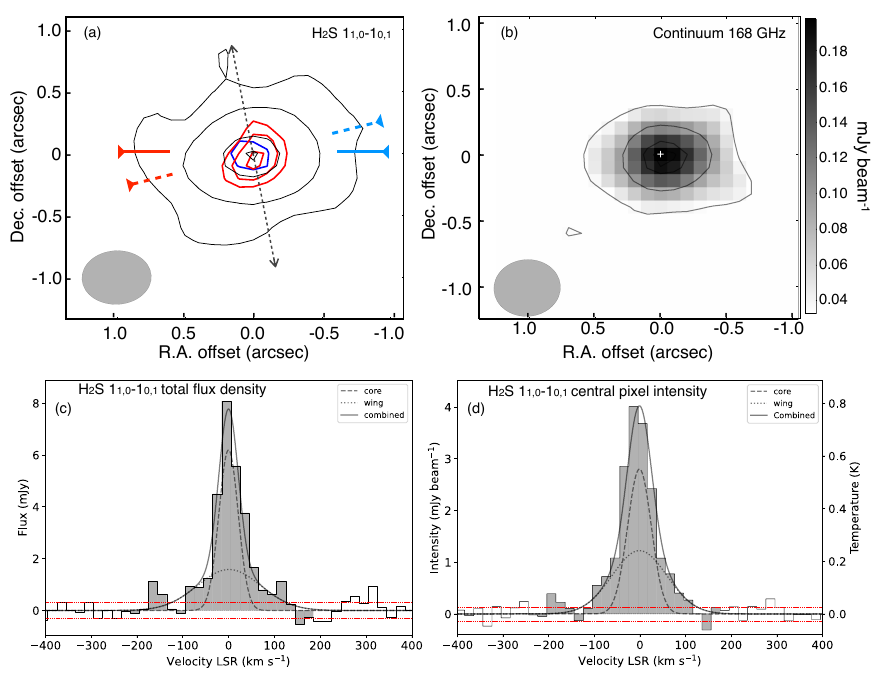}
    \caption{The results of 168~GHz observations towards NGC\,1377. 
    (a) The velocity-integrated line emission maps of H$_2$S 1$_{1,0}$-1$_{0,1}$: black contours: -70 km s$^{-1}$ to 70 km s$^{-1}$ (3,9,27, and 36 $\sigma$ where $\sigma_{\rm core} = 6.94 \times 10^{-3}$ Jy~beam$^{-1}$ km s$^{-1}$), blue: -200 km s$^{-1}$ to -100 km s$^{-1}$, red: 100 km s$^{-1}$ to 200 km s$^{-1}$ (3,4, and 6 $\sigma$ where $\sigma_{\rm red} = 9.35 \times 10^{-3}$ Jy~beam$^{-1}$ km s$^{-1}$ and $\sigma_{\rm blue} = 9.38 \times 10^{-3}$ Jy~beam$^{-1}$ km s$^{-1}$, respectively).  The beam size is shown at the bottom-left in grey-filled circle. 
    The blue and red line segments indicate the position angle (PA) of the disk. The PA of the nuclear continuum is 90$^{\circ}$ (scales of r=2 pc) and also the orientation of the dynamics of the nuclear disk is close to 90$^{\circ}$ \citep{Aalto2020ALMA1377}. \citet{Aalto2020ALMA1377} note that this is different from the 104$^{\circ}$ found for the major axis on larger scales of r=10 pc \citep{Aalto2016}. The direction of the molecular outflows detected in CO~3-2 \citep[PA $\sim 11^{\circ}$, Fig.\ref{fig:1377CO32}, ][]{Aalto2016} is indicated with black dotted line.
    (b) The Continuum emission at 168~GHz in greyscale and contours: (3,9,15, and 17 $\sigma$ where $\sigma_{\rm cont} = 1.1 \times 10^{-5}$ Jy~beam$^{-1}$).
    (c) Spectrum in the unit of the flux density within the emitting region (over 3 $\sigma$ in moment 0 image) against the line velocity offset at the systemic velocity. The black histogram shows the observation result. The red-dotted line indicates the 1 $\sigma$ rms level. The black dashed line:narrow line component, dotted line: broad line component, and solid line: two components combined.
    (d) Spectrum in unit of the mean brightness at the peak intensity pixel against the line velocity.}
    \label{fig:1377H2S11}
\end{figure*}

\begin{figure*}
    \centering
    \includegraphics[width=0.8\textwidth]{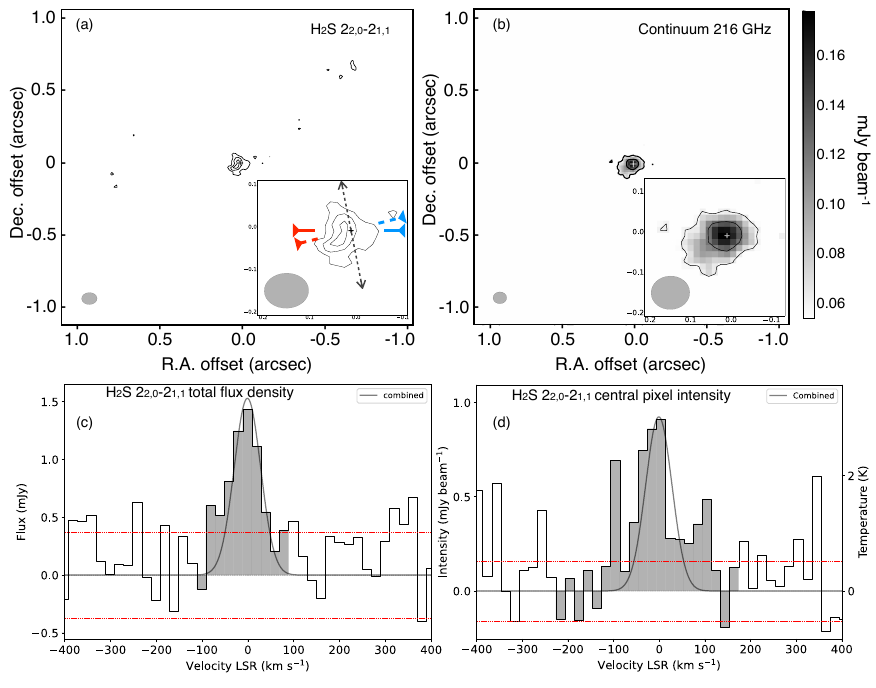}
    \caption{The results of 216~GHz observations towards NGC\,1377. (a) The velocity-integrated line emission maps of H$_2$S 2$_{2,0}$-2$_{1,1}$:  -100 km s$^{-1}$ to 100 km s$^{-1}$ (2, 3, 4, and 5 $\sigma$ where $\sigma = 2.1 \times 10^{-2}$ Jy~beam$^{-1}$ km s$^{-1}$). The lines and symbol conventions are the same as in Fig.\ref{fig:1377H2S11}.
    (b) The Continuum emission at 216~GHz in greyscale and contours: (3,6, and 9 $\sigma$ where $\sigma_{\rm cont} = 1.8 \times 10^{-5}$ Jy~beam$^{-1}$).
    (c) Spectrum in the unit of the flux density within the emitting region (over 3 $\sigma$ in moment 0 image) against the line velocity offset at the systemic velocity. The black histogram shows the observation result. The red-dotted line indicates the 1 $\sigma$ rms level. The black solid line: single Gaussian fit.
    (d) Spectrum in unit of the mean brightness at the peak intensity pixel against the line velocity.}
    \label{fig:1377H2S22}
\end{figure*}

\begin{figure*}
    \centering
    \includegraphics[width=0.8\textwidth]{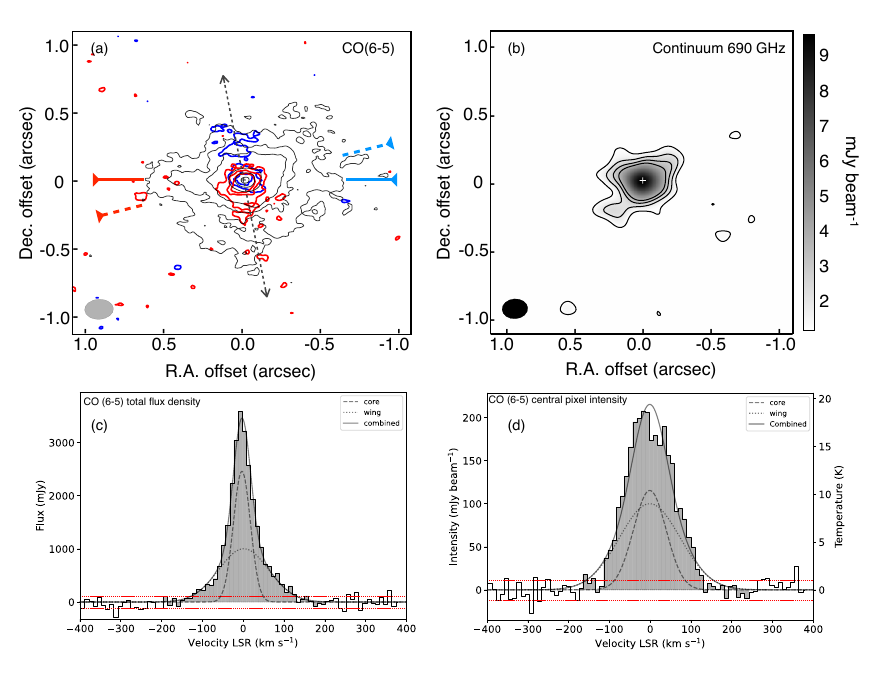}
    \caption{The results of 691~GHz observations towards NGC\,1377. (a) The velocity-integrated line emission maps of CO~6-5: black contours: -70 km s$^{-1}$ to 70 km s$^{-1}$ (3,6,9,27, and 36 $\sigma$ where $\sigma_{\rm core}$ = 0.29~ Jy~beam$^{-1}$ km s$^{-1}$), blue: -200 km s$^{-1}$ to -100 km s$^{-1}$, red: 100 km s$^{-1}$ to 200 km s$^{-1}$ (3,6,9, and 12 $\sigma$ where $\sigma_{\rm red}$ = 0.13~Jy~beam$^{-1}$ km s$^{-1}$ and $\sigma_{\rm blue}$ = 0.16~Jy~beam$^{-1}$ km s$^{-1}$, respectively). The lines and symbol conventions are the same as in Fig.\ref{fig:1377H2S11}.
    (b) The Continuum emission at 691~GHz in greyscale and contours: (3,5,7, and 9 $\sigma$ where $\sigma_{\rm cont} = 4.07 \times 10^{-1}$ Jy~beam$^{-1}$).
    (c) Spectrum in the unit of the flux density within the emitting region (over 3 $\sigma$ in moment 0 image) against the line velocity offset at the systemic velocity. The black histogram shows the observation result. The red-dotted line indicates the 1 $\sigma$ rms level. The black dashed line:narrow line component, dotted line: broad line component, and solid line: two components combined.
    (d) Spectrum in unit of the mean brightness at the peak intensity pixel against the line velocity.}
    \label{fig:1377CO}
\end{figure*}

\begin{figure}
    \centering\includegraphics[width=0.4\textwidth]{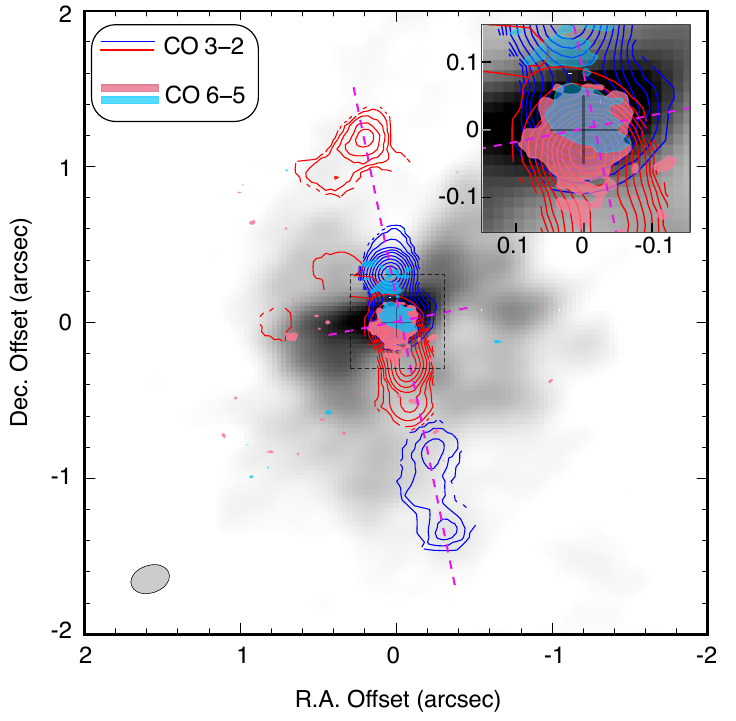}
    \caption{
    The velocity-integrated map of CO 3-2 \citep{Aalto2016}. Greyscale shows the emission close to systemic velocity. The high velocity  emission from the molecular jet is shown in contours (with the red and blue showing velocity reversals along the axis). The vertical bar indicates a scale of 100 pc. (For details on the figure see \citet{Aalto2016}). We have indicated the extent and orientation of the CO 6-5 high velocity gas (from Fig. \ref{fig:1377CO}) with light blue and yellow ovals.
    }
    \label{fig:1377CO32}
\end{figure}

%-----------end Figures : Spectra & Maps for three sources----------------------------------

\subsection{Radiative Transfer Modelling}
\label{subsec:rt}
We used the RADEX non-LTE radiative transfer code \citep{VanDerTak2007ARatios} to estimate the physical conditions ($n_{\rm H_2}$, $T_{\rm kin}$, $N_{\rm H_2S}$) of the gas. 
We used the collision rates of H$_2$S which are scaled from the full quantum calculations for collisional excitation of ortho- and para-H$_2$O by ortho- and para-H$_2$ from the references:\citet{Dubernet2006InfluenceTemperature,Dubernet2009RotationalTemperature, Daniel2010RotationalTemperature,Daniel2011RotationalCoefficients}.
 Additionally, we adopt an ortho-to-para ratio of 3 for H$_2$S (H$_2$S 1$_{1,0}$-1$_{0,1}$ is ortho, while H$_2$S 2$_{2,0}$-2$_{1,1}$ is para), corresponding to the thermal equilibrium value, and assume that H$_2$S is primarily excited through collisions \citep{Crockett2014}. 
 
Because each H$_2$S transition was observed with a slightly different beam, we adopt the smaller of the two synthesized beams as the effective source size. This allows us to assume that the measured intensities for both lines trace the same region and mitigates the effect of beam size mismatches.

 We explore molecular hydrogen densities $n_{\rm H_2}$ in the range $10^4$–$10^8$ cm$^{-3}$ (sampled in 40 logarithmically spaced steps) which brackets the critical densities $5.4 \times 10^5$ cm$^{-3}$ and $1 \times 10^6$ cm$^{-3}$  of the ortho- and para-H$_2$S transitions, respectively, at $T=100$~K \footnote{The critical densities $n_{\rm crit}$ [cm$^{3}$] are calculated as $n_{\rm crit} = \frac{A}{a_{\rm col}}$, where $A$ [s$^{-1}$] is the Einstein A-coefficient and $a_{\rm col}$ [cm$^{-3}$~s$^{-1}$] is the collisional rate coefficient. Values are from the molecular data from Leiden Atomic and Molecular Database (LAMDA:\url{https://home.strw.leidenuniv.nl/~moldata/}.)}. 
 The H$_2$S column density N$_{\rm H_2S}$ is varied between $10^{14}$ and $10^{18}$ cm$^{-2}$ (sampled in 40 logarithmically spaced steps), consistent with the possibility of very high H$_2$ columns ($10^{24}$ cm$^{-2}$ ) and a typical H$_2$S abundance of $10^{-10}$–$10^{-6}$ \citep{Holdship2017, Crockett2014, Sato2022APEXOutflows}. Finally, we consider kinetic temperatures $T_{\rm kin}$=40-200 K with 15 steps evenly spaced on a linear scale, constrained by the observed H$_2$S brightness temperatures and previous CO-based temperature estimates \citep{Aalto2016}.\\
 
%The RADEX model parameters are as follows: 
%The molecular hydrogen density ($n_{\rm H_2}$) is set within the range %$10^4$–$10^8$ cm$^{-3}$, chosen to be approximately within two orders %of magnitude of the critical densities of the observed ortho- and para-%H$_2$S transitions, which are $5 \times 10^5$ cm$^{-3}$ and $1 \times %10^6$ cm$^{-3}$ at $T_{\rm kin} = 100$ K (see Appendix for details).  
%The H$_2$S column density ($N_{\rm H_2S}$) is varied between $10^{14}$ %and $10^{18}$ cm$^{-2}$, based on estimates that the H$_2$ column %density in the nuclear regions could be as high as $10^{24}$ cm$^{-2}$ %\citep{Aalto2016}, and the expected ortho-H$_2$S abundance of %$10^{-10}$–$10^{-6}$ \citep{Holdship2017, Crockett2014, %Sato2022APEXOutflows}.  
%The kinetic temperature ($T_{\rm kin}$) is set between 40 and 200 K. %The lower limit is determined by the brightness temperatures of the %H$_2$S 1$_{1,0}$-1$_{0,1}$ transition, while the upper limit is based %on estimates from CO observations of NGC\,1377 \citep{Aalto2016}.  

We compared the RADEX predictions to two observationally derived ratios for each source: 

\begin{enumerate}
\item The peak brightness temperature ratio, $T_{{\rm b},22}$/$T_{{\rm b},11}$, measured at the central pixel of the data cube. \,

\item The integrated brightness temperature ratio, $\int{T_{{\rm b},22} dv}/\int{T_{{\rm b},11}} dv$, derived from the total velocity-integrated 
intensities. 

\end{enumerate}

\noindent
The first ratio constrains the excitation under near-peak conditions, while the second ratio is more sensitive to the total column of emitting gas over the full line profile. We searched for model solutions that reproduce the observed ratio between these two quantities.

%To compare the RADEX models with the observations, we use two quantities from the two lines: the ratio of the brightness temperatures ($T_{\rm b}$ [K]) derived from the peak intensity of the central pixel and the ratio of the velocity-integrated brightness temperatures ($T_{\rm b} dv$ [K~km~s$^{-1}$]) obtained from the velocity-integrated intensity. We searched for model solutions that reproduce the observed ratio between these two quantities.

%For the velocity width in the RADEX models, we adopted the average FWHM of the core components of the two transitions. The parameters used in the modelling are summarised in Table \ref{tab:Tb}.

We used the average FWHM of each transition’s core component as the velocity width input to RADEX, thereby excluding broad wings that may arise from outflows (separate features in the fitting). We list all final modelling parameters, along with their assumed ranges, in Table \ref{tab:Tb}. Uncertainties may arise if any significant fraction of the line emission (especially in the wings) originates from gas with different physical conditions. \\
%but our primary goal is to characterize the densest core regions.

\noindent
Although this approach provides robust first-order constraints, we note several caveats: (a) partial beam dilution could reduce our line ratios if the true source size is smaller than the assumed beam, (b) scaling collision rates from H$_2$O introduces additional uncertainties for H$_2$S, (c) the assumed ortho-to-para (o/p) ratio of 3 may not apply to regions with non-thermal processes. (For example, in the Orion KL region, \citet{Crockett2014} calculated an o/p ratio of less than 2.), and (d) the RADEX modelling is constrained by only two H$_2$S transitions.  Despite these limitations, the models give a consistent picture of dense ($>10^6$ cm$^{-3}$) and warm ($T_{\rm kin}>$ 40 K) gas traced by H$_2$S. \\

%%% Explain about CO for N1377
%The same method was applied to determine the physical conditions of CO, using the $T_b$ ratio of CO 3-2 to CO 6-5, for NGC\,1377.\\

%%%-------------------------------------
% When the LAMDA molecule data was used
%%%-------------------------------------

% This should be used in the arguments somewhere.%%%
% We estimate the collision rates of H\(_2\)S with ortho and para-H\(_2\) by scaling the existing H\(_2\)O rates from Dubernet et al (2009), (Daniel/Dubernet et al 2012), and Daniel et al. (2011). This scaling ensures that the para-H\(_2\) rate (similar to He as a collision partner) for the 1\(_{1,0}\)–1\(_{0,1}\) transition aligns with experimental results at low temperatures. To achieve this consistency, we applied a constant factor of 0.4 to the H\(_2\)O rates. We selected H\(_2\)O as an analogue for H\(_2\)S due to the similarity in their molecular structures.\\
% Infrared pumping to the higher transition might affect the population of lower transitions of H\(_2\)S. Though \citet{Crockett2014} reported that the effect on those transitions we used should not be significant.
%If the real source size is even smaller than the beam sizes of the observations, the true brightness temperature could be higher.\\

\subsubsection{Results for individual objects}
\label{subsubsec:rt_individual}
\textbf{NGC\,1377}:  
Using a line width \(\Delta v = 65\) km s\(^{-1}\), we find that an H\(_2\) density \(n_{\rm H_2} > 3 \times 10^6\) cm\(^{-3}\), at any H\(_2\)S column density \(N_{\rm H_2S}\) and kinetic temperature \(T_{\rm kin}\), can adequately reproduce our observational results. Since this value is well above the critical density for both transitions, the gas traced by H\(_2\)S is most likely thermalised.  
The derived density is at least ten times higher than that obtained from CO by \citet{Aalto2020ALMA1377}, who reported a column density of \(N_{\rm H_2} = 1.8 \times 10^{24}\) cm\(^{-2}\) in the inner 2 pc region (beam size = \(0''.02\)), corresponding to \(n_{\rm H_2} = 1 \times 10^5\) cm\(^{-3}\).  

\begin{figure*}
    \centering
    \includegraphics[width=\textwidth]{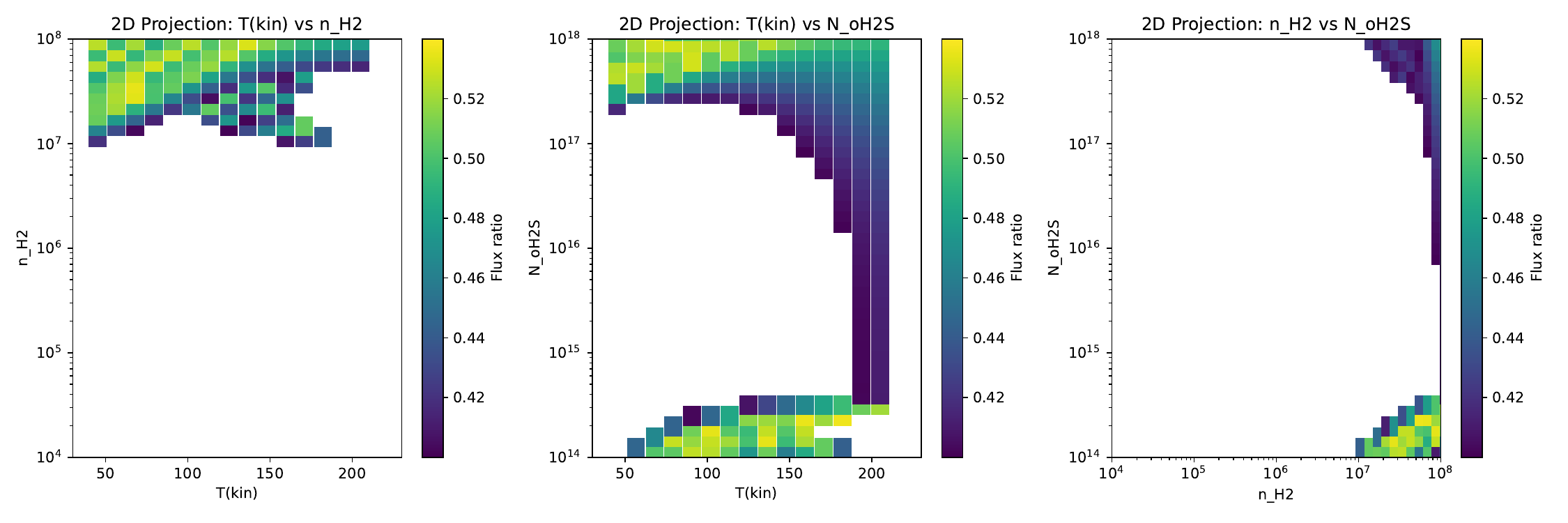}
    \caption{The result from RADEX modelling for NGC\,4418. The line ratio of H$_2$S 2$_{2,0}$-2$_{1,1}$ to H$_2$S 1$_{1,0}$-1$_{0,1}$ (in colour) is plotted in the parameter spaces of \textit{left}: $n_{\rm H_2}$ vs T$_{\rm kin}$, \textit{middle}: $N_{\rm H_2S}$ vs T$_{\rm kin}$, and \textit{right}: $N_{\rm H_2S}$ vs $n_{\rm H_2}$.}
    \label{fig:rt}
\end{figure*}

\textbf{NGC\,4418}:  
With a line width \(\Delta v = 108\) km s\(^{-1}\), we find that an H\(_2\) density \(n_{\rm H_2} > 1 \times 10^7\) cm\(^{-3}\), at any H\(_2\)S column density \(N_{\rm H_2S}\) and kinetic temperature \(T_{\rm kin}\), can reproduce our observational results (Fig. \ref{fig:rt}). Thus we expect that H$_2$S traces the gas with an even higher \(n_{\rm H_2}\) than those in NGC\,1377.  
This inferred density exceeds previous estimates from CO observations. 
Prior studies suggest a stratified temperature and density structure, with a 300–500 K layer in the nuclear region (5–8 pc) and a surrounding 160 K layer extending to \(\sim40\) pc \citep{Costagliola2013, Costagliola2015, Sakamoto2021DeeplyALMA}. 
Our modelling can not distinguish those two layers due to the limitation of the spatial resolutions.

\textbf{NGC\,1266}:  
For a line width \(\Delta v = 56\) km s\(^{-1}\), we find that an H\(_2\) density \(n_{\rm H_2} > 5 \times 10^7\) cm\(^{-3}\), a kinetic temperature \(T_{\rm kin} > 140\) K, and an H\(_2\)S column density \(N_{\rm H_2S} = 1 \times 10^{14}\) cm\(^{-2}\) can reproduce our observational results. Notably, the modelling solutions were found only at this specific H$_2$S column density.  
Our results indicate an extremely high \(n_{\rm H_2}\), consistent with a highly dense environment. However, it should be noted that these values are averaged over the beam size, covering a region of \(\sim75\) pc, potentially blending gas components with different densities and temperatures.  
For comparison, \citet{Alatalo2011} estimated a lower limit of \(n_{\rm H_2} \geq 6.9 \times 10^{3}\) cm\(^{-3}\) from CO 1-0 observations of the central region with a radius of \(\sim1''\) (\(\sim150\) pc). The significant difference in derived densities likely reflects the use of different tracers and the higher critical density of H\(_2\)S, which allows us to probe denser regions within the beam.  

% Caveats
%It is important to emphasise that the values obtained from our RADEX modelling should be interpreted with caution, as they depend on specific assumptions about the input parameters.
%Firstly, the compared brightness temperatures are beam-size corrected under the reasonable assumption that the actual source size is comparable to the smallest synthesised beam of the observations of the two H$_2$S transitions.
%If this assumption does not hold, the derived line ratios would change, leading to different estimates of density and temperature.
%Secondly, we assume that the cloud is in thermal equilibrium, which directly determines the ortho-to-para (o/p) ratio of H$_2$S.
%If the gas is sub-thermal, the o/p ratio could deviate from the assumed value of 3 \citep{Crockett2014}. In the Orion KL region, \citet{Crockett2014} calculated an o/p ratio of less than 2. 
%If the ratio is closer to unity, [explain the consequences there. how it affects the derived physical conditions]. 
%Additionally, the RADEX modelling is constrained by only two observed H$_2$S transitions, and the collision rates used were scaled from those of H$_2$O. 
%These limitations may introduce uncertainties in the derived physical parameters, meaning that the results should be interpreted with care when drawing conclusions.

%-----------begin Figure "RADEX result N4418"------------------------------------------------
%-----------end Figure "RADEX result N4418"------------------------------------------------

%-----------begin Table "Observations results"------------------------------------------------
\begin{table*}[!h]
	\caption{Observations results}             % title of Table
	\label{tab:res2}      % is used to refer this table in the text
	\centering                          % used for centering table
	\begin{tabular}{l c c c c  @{\hskip 2pt}c c c c c}        % centered columns (4 columns)
		\hline\hline                 % inserts double horizontal lines
 & \multicolumn{4}{c}{Central beam} & & \multicolumn{4}{c}{3 $\sigma$ region}\\
  \cline{2-5} \cline{7-10}
 & FWHM  & $v_{\rm centre}$ & $I_{\rm peak}$  & $\int S dv$ \tablefootmark{1}& & FWHM  & $v_{\rm centre}$ & $I_{\rm peak}$  & $\int S dv$ \tablefootmark{1}\\
 & [km s$^{-1}$] &  [km s$^{-1}$] & [mJy beam$^{-1}$] &  [Jy beam$^{-1}$ km s$^{-1}$] & & [km s$^{-1}$] &  [km s$^{-1}$] & [mJy] &  [Jy km s$^{-1}$] \\
 \hline
\textit{NGC\,1377} \\
%\vspace{1pt}\\
\hline
H$_2$S 1-1
Total & 80$\pm$3 & 3$\pm$1 & 3.9$\pm$0.1 & 0.36$\pm$0.01 && 67$\pm$4  & 1$\pm$2 & 7.3$\pm$0.4 & 0.52$\pm$0.04 \\ % from 1 comp fitting @N1377_line_fitting.ipynb
Core & 57$\pm$10 & 1$\pm$2 & 2.7$\pm$0.7 & 0.17$\pm$0.05 && 50$\pm$6 & 0$\pm$2 & 6.2$\pm$0.8 & 0.33$\pm$0.06  \\
Wings & 129$\pm$28 & 7$\pm$7 & 1.4$\pm$0.8 & 0.20$\pm$0.10 && 163$\pm$45 & 1$\pm$12 & 1.6$\pm$0.7 & 0.27$\pm$0.15 \\
\hline
H$_2$S 2-2
%Total & 152$\pm$11 & -10$\pm$4 & 2.4$\pm$0.1 & 0.40$\pm$0.02 && 136$\pm$10 & -4$\pm$4 & 3.0$\pm$0.2 & 0.44$\pm$0.04 \\
Core & 68$\pm$36 & -1$\pm$5 & 0.9$\pm$0.4 & 0.07$\pm$0.05 && 68$\pm$27 & -1$\pm$12 & 1.5$\pm$0.5 & 0.1$\pm$0.06 \\
\hline
CO(6-5)
Total & 138$\pm$3 & -2$\pm$1 & 207$\pm$4 & 30$\pm$1 && 77$\pm$2 & -3$\pm$1 & 2.9$\pm$0.1 Jy & 239$\pm$9  \\
Core & 94$\pm$25 & 0$\pm$3 & 115$\pm$71 & 11.7$\pm$7.8 && 50$\pm$3 & -3$\pm$1 & 2.5$\pm$0.1 & 129$\pm$10 \\
Wings & 175$\pm$39 & 0$\pm$6 & 100$\pm$73 & 18.6$\pm$14.2 && 144$\pm$11 & 1$\pm$3 & 1.0$\pm$0.1 & 154$\pm$26 \\
\hline
\textit{NGC\,4418} \\
\hline
H$_2$S 1-1 
Total & 121$\pm$4 & -5$\pm$2 & 29$\pm$1 & 4.1$\pm$0.1&& 134$\pm$5 & -4$\pm$2 & 62$\pm$2 & 9.6$\pm$0.3 \\
Core & 107$\pm$4 & -1$\pm$1 & 27$\pm$1& 3.0$\pm$0.1 && 115$\pm$5 & -7$\pm$1 & 56$\pm$2 & 6.8$\pm$0.4\\
Wings & 353$\pm$60 & -1$\pm$16 & 2.9$\pm$0.8 & 1.1$\pm$0.4 && 312$\pm$40 & 18$\pm$26 & 8.6$\pm$2.2 & 2.8$\pm$0.8  \\
\hline
H$_2$S 2-2 
Total & 155$\pm$11 & -6$\pm$5 & 22$\pm$1 & 4.1$\pm$0.1  && 141$\pm$11 & -5$\pm$5 & 50$\pm$3 & 8.2$\pm$0.2 \\
Core & 109$\pm$18 & -5$\pm$6 & 17$\pm$5 & 2.0$\pm$0.7  && 118$\pm$22 & -2$\pm$6 & 47$\pm$7 & 5.9$\pm$2.4  \\
Wings& 236$\pm$59 & 8$\pm$56 & 6.1$\pm$4.9 & 1.5$\pm$1.3  && 283$\pm$26 & -5$\pm$11 & 5$\pm$18 & 1.5$\pm$3.1  \\
Other& 99$\pm$42 & -177$\pm$20 & 4.2$\pm$2.2 & 0.4$\pm$0.3  && 107$\pm$44 & -166$\pm$21 & 9.9$\pm$2.3 & 1.1$\pm$0.5  \\
\hline
\textit{NGC\,1266} \\
\hline
H$_2$S 1-1
Total & 98$\pm$5 & -18$\pm$2 & 4.2$\pm$0.2 & 0.43$\pm$0.02 &&  131$\pm$8 & 19$\pm$4 & 9.9$\pm$0.2 & 1.4$\pm$0.1  \\
Core & 71$\pm$17 & -1$\pm$5 & 3.0$\pm$1.0 & 0.22$\pm$0.09 && 71$\pm$39 & 1$\pm$10 & 5.7$\pm$5.3 & 0.4$\pm$0.5  \\
Wings & 188$\pm$84 & -1$\pm$52 & 1.4$\pm$1.0 & 0.28$\pm$0.24 && 236$\pm$217 & -1$\pm$247 & 5.0$\pm$8.5 & 1.3$\pm$2.4  \\
Other &  118$\pm$87 & -156$\pm$54 & -0.7$\pm$0.6 & -0.09$\pm$0.1 && 140$\pm$214 & -144$\pm$125 & -3.2$\pm$12.6 & -0.5$\pm$2.0  \\
\hline
H$_2$S 2-2 
Core & 51$\pm$10 & -1$\pm$4 & 6.1$\pm$1.0 & 0.33$\pm$0.09 && 59$\pm$12 & 0$\pm$5 & 4.7$\pm$0.9 & 0.29$\pm$0.08 \\
\hline\hline
	\end{tabular}
	\tablefoot{
		\tablefoottext{1}{The velocity integrated flux of the whole line is obtained by integrating the spatially integrated flux of each channel.};
	}
\end{table*}

\begin{table*}[h]
\caption{Brightness temperatures}                 % title of Table
\label{tab:Tb}    % is used to refer this table in the text
\centering                        % used for centering table
\begin{tabular}{c c c c c c c c}      % centered columns (4 columns)
\hline\hline               % inserts double horizontal lines
Source & $\int T_{\rm b11} dv$ \tablefootmark{a} & $\int T_{\rm b22} dv$ \tablefootmark{b}  & $R_{\rm Tb22dv/Tb11dv}$ \tablefootmark{c} &$T_{\rm b11}$ & $T_{\rm b22}$ & $R_{\rm Tb22/Tb11}$ & $\Delta v$ \tablefootmark{d} \\         % table heading
 & [K~km~s$^{-1}$] & [K~km~s$^{-1}$] &  &   [K] & [K] && [km~s$^{-1}$] \\
\hline                      % inserts single horizontal line
   NGC1377 & 9.3$\times$10$^2$ & 2.4$\times$10$^2$ & 0.25 & 25 & 7 & 0.28 & 65 \\    % inserting body of the table
   NGC4418 & 3.9$\times$10$^3$ & 1.6$\times$10$^3$ & 0.40 &  41 & 22 & 0.54 & 108 \\
   NGC1266 & 32 & 40 & 1.25 & 3.3 & 3.7 & 1.12 & 56 \\
\hline                                  %inserts single line
\end{tabular}
\tablefoot{This table summarises the values from the observations which are referenced for the RADEX modelling.\\
\tablefoottext{a}{The velocity-integrated brightness temperatures of H$_2$S 1$_{1,0}$-1$_{0,1}$ transition.}
\tablefoottext{b}{The velocity-integrated brightness temperatures of H$_2$S 2$_{2,0}$-2$_{1,1}$ transition.}
\tablefoottext{c}{The ratios of (a) to (b).}
\tablefoottext{d}{The FWHMs of each line, from a Gaussian fit.}
}
\end{table*}
%%%%%%%%%%%%%%%%%%%%%%%%%%%%%%%%%%%%%%%%%%%%%%%%%%%%%%%%%%%%%%

\subsection{Continuum emission}
\label{subsec:cont}
% General info about shown continuum results.
The continuum emission maps are presented in the panel (b) of Figs. \ref{fig:1377H2S11} to \ref{fig:1266H2S22}.
Continuum emission was detected for all three sources.
For the measurements of the total flux and the spatial extent of the emission we included only signal above 3 $\sigma$ flux levels.

% Continuum info for each galaxy
%N1377
The size of the 168~GHz continuum emission in NGC\,1377 is approximately $1.3'' \times 1.0''$ ($133 \times 102$ pc). The peak flux of this emission is 0.20~mJy~beam$^{-1}$ ($T_b$=1.5~K), and the total flux, integrated over the region defined by the 3 $\sigma$ contour level, is $0.30 \pm 0.01$ mJy.
%The peak flux of this emission is 0.198~mJy~beam$^{-1}$ ($T_b$=1.5~K), and the total flux, integrated over the region defined by the 3 $\sigma$ contour level, is $0.30 \pm 0.013$ mJy.
The 216~GHz continuum emission in NGC\,1377 arises from a more compact region with an angular size of approximately $0.2''$ in diameter, corresponding to only 20 pc across the emitting region.
The peak flux of this emission is 0.18~mJy~beam$^{-1}$ ($T_b$=3.5~K), and the total flux, integrated over the region defined by the 3 $\sigma$ contour level, is $0.28 \pm 0.02$ mJy.
%The peak flux of this emission is 0.178~mJy~beam$^{-1}$ ($T_b$=3.5~K), and the total flux, integrated over the region defined by the 3 $\sigma$ contour level, is $0.28 \pm 0.02$ mJy.
The 691~GHz continuum emitting region has a diameter of $0.8''$ (82 pc), with significantly higher total and peak fluxes of $32 \pm 2$ mJy and 9.7~mJy~beam$^{-1}$ ($T_b$=14~K), respectively, compared to those at 168~GHz and 216~GHz.
%The 691~GHz continuum emitting region has a diameter of $0.8''$ (82 pc), with significantly higher total and peak fluxes of $31.5 \pm 2.4$ mJy and 9.65~mJy~beam$^{-1}$ ($T_b$=14~K), respectively, compared to those at 168~GHz and 216~GHz.
As shown in Figs.\ref{fig:1377H2S11}, \ref{fig:1377H2S22} and \ref{fig:1377CO}, while the 691~GHz continuum emission is spatially resolved by the synthesised beam, the emission at 168~GHz and 216~GHz is not resolved.

%N4418
NGC 4418 exhibits similar sized continuum regions but higher fluxes at both measured frequencies compared to the continuum emission of NGC 1377. 
The 168~GHz continuum emitting region in NGC\,4418 has a diameter of approximately $0.5''$ (80 pc). 
The peak flux of this emission is 18~mJy~beam$^{-1}$ ($T_b$=28~K), and the total flux, integrated over the region defined by the 3 $\sigma$ contour level, is $23.4 \pm 0.2$ mJy.
% The peak flux of this emission is 18.4~mJy~beam$^{-1}$ ($T_b$=28~K), and the total flux, integrated over the region defined by the 3 $\sigma$ contour level, is $23.4 \pm 0.15$ mJy.
The size of the 216~GHz continuum emission in NGC\,4418 has an angular size of $1.0''$ (165 pc).
The peak flux of this emission is 24.1 mJy~beam$^{-1}$ ($T_b$=12~K) and the total flux is $45.0 \pm 1.2$ mJy.
% The peak flux of this emission is 24.1 mJy~beam$^{-1}$ ($T_b$=12~K) and the total flux is $45.0 \pm 1.18$ mJy. The 216~GHz continuum emission is spatially resolved by the synthesised beam, whilst the emission at 168~GHz is not resolved (Fig.\ref{fig:4418}).

%N1266
NGC 1266 has the largest continuum emitting region among the three galaxies, with a diameter of 290 pc at both frequencies. 
The peak flux of the 168~GHz continuum emission is 2.6 mJy~beam$^{-1}$ ($T_b$=2.8~K)  and the total flux is $3.8 \pm 0.1$ mJy. 
%The peak flux of the 168~GHz continuum emission is 2.59 mJy~beam$^{-1}$ ($T_b$=2.8~K)  and the total flux is $3.84 \pm 0.11$ mJy. 
The 216~GHz continuum emission within the 3 $\sigma$ contour region has a peak flux of 5.2 mJy~beam$^{-1}$ ($T_b$=2.4~K) and the total flux is $5.7 \pm 0.2$ mJy.
% The 216~GHz continuum emission within the 3 $\sigma$ contour region has the peak flux of 5.22 mJy~beam$^{-1}$ ($T_b$=2.4~K) and the total flux is $5.74 \pm 0.24$ mJy.

%%%%%%%%%%%%%%%%%%%%%%%%%%%%%%%%%%%%%%%%%%%%%%%%%%%%%%%%%%%%%%
\section{Discussions}
\label{sec:discu}
In a recent study, \citet{Sato2022APEXOutflows} presented single-dish observations toward a group of 12 galaxies, revealing the presence of H$_2$S emission in 9 of them. Although the angular resolution of our observations ($\sim$ 37") did not allow us to resolve the emitting region, the lack of correlation between the normalised intensity of the H$_2$S lines and the presence of observed molecular outflows led us to conclude that the H$_2$S abundance enhancement does not seem to be directly linked to galactic-scale outflows. Instead, for certain galaxies, including NGC\,4418, we proposed that the enhancement could be attributed to radiative processes or small-scale shocks. Additionally, \textit{a correlation between H$_2$S line luminosity and outflow mass hinted at a connection between the dense gas reservoir and the evolution of molecular feedback.} The much higher angular resolution ALMA observations (0".1 - 1".2) presented in this work have now provided further insights on the origin of the H$_2$S in these galaxies. Our new ALMA observations resolve these regions down to scales of 20–30 pc, revealing compact H$_2$S emission at galaxy centres.
%The new data reveal that the H$_2$S emission is compact, concentrated %within regions of approximately $\lesssim$ 20–30 pc in size, toward %the centre of the galaxy. 
Additionally, some galaxies exhibit relatively broad spectral line wings, hinting at a connection \textit{between the H$_2$S emission and the presence of a high-velocity outflow.} 

However, a puzzling discrepancy arises: the H$_2$S-emitting gas appears significantly denser (\(10^7\)-\(10^8\) cm\(^{-3}\)) than the density inferred from CO emission (\(\sim10^5\) cm\(^{-3}\)) in NGC\,1377 (Sec.~\ref{subsec:density}).
This difference is worth noting given that the CO-emitting region, from which the density was obtained, is only 2~pc across.
It is important to note, though, that the CO emission does not necessarily originate from the same 2~pc region as the H$_2$S emission. 
In order to understand the results from our observations and modelling, in the following we consider some likely scenarios for the origin of H$_2$S emission in these galaxies.

\subsection{Possible Mechanisms for the Formation of Gas-Phase H$_2$S} % radiation? shock? sputtered?
\label{subsec:formh2s}
The detection of H$_2$S in the gas phase implies that one or more mechanisms must exist to release this molecule from dust grains, where it is expected to predominantly form \citep{Charnley1997,Wakelam2004,Viti2004}. These mechanisms can enhance the gas-phase abundance of H$_2$S through several processes. The high density of the gas alone cannot account for the enhanced H$_2$S abundance, as chemical desorption becomes inefficient at densities above $2 \times 10^4$ cm$^{-3}$
 under dark (high visual extinction) and cold ($\sim$ 10~K) conditions \citep[e.g.][]{Vidal2017, Navarro-Almaida2020GasCase}. This suggests that additional processes must be at play to sustain a sufficient amount of gas-phase H$_2$S for emission. 

Radiation from a nuclear starburst or an active galactic nucleus (AGN) can lead to the sublimation of volatile species such as H$_2$S through thermal desorption, efficiently releasing molecules into the gas phase as a result of increased dust temperatures \citep[e.g.][]{MInh1990,Charnley1997,Bachiller2001Chemically1157, Hatchell2002PossibleCores, Minh2007, Woods2015}. Alternatively, in even lower-temperature environments, photo-desorption could also contribute to molecule release, particularly if the dust temperature is around 10–50 K \citep{Goicoechea2021BottlenecksGrains}. Thermal desorption can also be triggered by shocks. The passage of a shock wave through the interstellar medium can compress and heat both the gas and dust, leading to the sublimation of H$_2$S from grain mantles. If the shock is relatively mild, such as a C-type shock, it can raise the gas temperature without completely dissociating H$_2$S molecules \citep[e.g.][]{PineaudesForets1993, Holdship2017}. Evidence for the potential role of this mechanism can be provided by the presence of broad line wings in the spectral lines of this molecule.

In regions with high-velocity shocks, direct sputtering may also be a viable mechanism. In this process, energetic particles or ions impact dust grains, physically ejecting H$_2$S into the gas phase. However, if the shock velocity is too high (e.g., J-type shocks exceeding $\sim$ 80 km s$^{-1}$), it could lead to the complete destruction of H$_2$S molecules, reducing rather than enhancing their abundance \citep[e.g.][]{Neufeld1989FASTSHOCK}.

To better understand the origin of the detected H$_2$S emission, it is crucial to determine both the kinetic temperature and the density of the gas in the observed regions. These physical conditions will help identify which of the above-mentioned mechanisms is most likely playing a major role in the production of gas-phase H$_2$S in the sample galaxies. A first hint at the origin of H$_2$S in the galaxies targeted in this work comes from the gas density derived from our RADEX modelling.

The processes discussed in this section, especially shock-induced desorption, offer a viable explanation for elevated H$_2$S in galaxy nuclei and may help resolve the puzzling density discrepancy in NGC1377, discussed in Sec.~\ref{subsec:density}.

\subsection{Origin of H$_2$S in the observed galaxies} % outflows? AGN rad? SB?
\label{subsec:orgh2s}
 As mentioned above, we find that a rather high H$_2$ density of \( n_{\mathrm{H}_2} > 3 \times 10^6 \) cm\(^{-3}\) can adequately reproduce our observational results. In particular, for NGC 4418 and NGC 1266, the implied density is \( >10^7 \) cm\(^{-3}\). Our modelling does not allow us to constrain the kinetic temperature within a range of 40–200 K, nor can the H$_2$S column density be adequately determined, except for NGC 1266, where the modelling yields solutions only for \( N_{\mathrm{H}_2S} = 1 \times 10^{14} \) cm\(^{-2}\).

A possible explanation for the high values of the density is that the H$_2$S emission originates from a compact structure within the central region of the galaxy whose size is smaller than the size probed by the synthesised beam of our observations (20-30~pc). %Thus, the actual regions from where the emission originates is highly concentrated, dense structure.
%potentially arising from a dense, accreting structure near a supermassive black hole (SMBH). In this scenario, the conditions of the gas accreting toward the nucleus could provide the necessary environment for H$_2$S formation, likely through thermal desorption driven by AGN radiation. 
The line profiles provide further clues to the origin of this emission. In particular, the line wings of the H$_2$S line may trace the base of an outflow. In NGC 1377, for example, the observed density gradient in the outflow (see Fig. \ref{fig:1377CO32}) is consistent with this scenario. Although Fig. \ref{fig:1377H2S11}(a) hints at a potential north-south shift in high-velocity gas, the magnitude of this shift is too small, relative to the beam size, to be claimed confidently. The wing emission could also emerge from unresolved Keplerian rotation close to the SMBH. Notably, higher-resolution observations of CO 6–5 in the nucleus of NGC 1377 reveal similar line wings, with dynamics that are not consistent with simple rotation \citep{Aalto2017Luminous1377}. This underscores the need for higher-resolution studies to clarify the structure and orientation of the H$_2$S line wing components.

Alternatively, the observed emission may arise from multiple compact, dense features distributed within the beam, such as shocked gas regions from molecular outflows or dense clumps associated with star formation.  \citet{Varenius2014TheRevealed} identified multiple compact radio sources in NGC\,4418, each smaller than 8 pc, within a 41 pc region, indicating intense star formation.

We emphasize the presence of line wings in several of the galaxies suggesting a link between the outflow and shocks. 
We also note the relatively short timescales ($\leq$~10$^4$~yr) over which H$_2$S can remain in the gas phase at high density \citep[e.g.][]{PineaudesForets1993}. 

If radiative pumping contributes significantly to excitation, our RADEX-derived densities may be overestimated, since the models assume purely collisional excitation.
%A third possibility is that, since our RADEX modelling does not include infrared background radiation, it may overestimate the gas density, and the observed line ratios could result from radiative pumping by infrared radiation rather than extreme gas densities. 
However, although H$_2$S can be radiatively pumped at densities exceeding \(10^7\) cm\(^{-3}\), lower excitation levels of H$_2$S, such as those in our study, appear to be less affected by this mechanism, as found by \citet{Crockett2014} in Orion KL.

Finally, NGC\,4418 presents a particularly intriguing case: the velocity gradient of the H$_2$S 2$_{2,0}$-2$_{1,1}$ emission appears to align with the large-scale stellar disk rotation observed by e.g. \citet{Wethers2024Double4418}, while the H$_2$S $1_{1,0} - 1_{0,1}$ line has an opposite orientation.  The inner gas disk has previously been reported to counterrotate with respect to the stars \citep[e.g.][]{Ohyama2019, Sakamoto2021DeeplyALMA}, and the H$_2$S $1_{1,0} - 1_{0,1}$ emission follows this general counterrotation. However, the nuclear gaseous dynamics is also found to be quite complex with multiple components. Red-shifted absorption structures are for example suggested to indicate an inflow of gas to the central region \citep[e.g.][]{Gonzalez-Alfonso2012,Costagliola2013, Sakamoto2013}. The H$_2$S $1_{1,0} - 1_{0,1}$ emission shows a red-shifted extension on the blue side of the nuclear rotation (Fig.~\ref{fig:4418H2S11}) which may signify gas inflowing in the disk, or a bar. Although this is also the same orientation as the larger-scale red-shifted outflow found by \citet{Wethers2024Double4418} with MUSE.  \citet{Sakamoto2021DeeplyALMA} discuss the possibility that a red-shifted absorption could be caused by a slanted outflow that is not parallel to the minor axis of the disk. The high velocity components of H$_2$S $1_{1,0} - 1_{0,1}$ show slightly extended emission along the minor axis of the disk (Red-shifted emission toward the north-west, and the blue-shifted emission toward the south-east: Fig.\ref{fig:4418H2S11}(a)).
Future, higher resolution observations are critical to determine if the redshifted H$_2$S $1_{1,0} - 1_{0,1}$ emission is due to inflowing gas, or a tilted outflow.

Why the H$_2$S $1_{1,0} - 1_{0,1}$ and 2$_{2,0}$-2$_{1,1}$ emission appear not to follow the same orientation is unclear, but may be caused by excitation effects in the complex nuclear dynamics. Furthermore, the blue-shifted wing of the H$_2$S 2$_{2,0}$-2$_{1,1}$ line is blended with other spectral features (see Fig.\,\ref{fig:4418H2S22}), making it difficult to derive a clear velocity gradient.

%This discrepancy suggests that the H$_2$S $1_{1,0} - 1_{0,1}$ line may %trace counter-rotation \citep[e.g.][]{Ohyama2019, Sakamoto2021DeeplyALMA} %or inflow. However, it should be noted that the blue-shifted wing of the %H$_2$S 2$_{2,0}$-2$_{1,1}$ line is blended with other spectral feature %(see Fig.\,\ref{fig:4418H2S22}), which makes difficult to derive clear %velocity gradient. Higher-resolution observations will be critical to %unravel the origin of this kinematic feature.}

\subsection{Why CO and H$_2$S yield different densities}
\label{subsec:density}

Interestingly, as mentioned above, the density derived from the CO observations within a compact region of 2~pc in NGC\,1377 is only \(\sim10^5\) cm\(^{-3}\), which is significantly smaller than the density derived from the H$_2$S. Obtaining different densities as derived from H$_2$S and CO observations is not uncommon. For example, \citet{Bouvier2024An253} find gas temperature of 30–159~K and \(n_{\rm gas} \geq 10^7\) cm\(^{-3}\) at the inner Circum Molecular Zone (CMZ) of NGC\,253, while from CO observations, the implied density is  \( 10^{3-4}\) cm\(^{-3}\) \citep{Tanaka2024VolumeAnalysis}. This difference has also been observed in other astrophysical environments, such as evolved stars. \citet{Gold2024Expanding2} recently reported the first detection of H$_2$S in the planetary nebula M1-59. These authors, in line with our approach, used RADEX modelling and found densities for H$_2$S greater than \(10^7\) cm\(^{-3}\), while the CO-derived densities were around \(10^4\) cm\(^{-3}\). Additionally, in another galactic example, ALMA observations of the evolved star W43A \citep{Tafoya2020ShapingJet} revealed that CO emission traces a collimated bipolar outflow, surrounded by higher-density material traced by dust and H$_2$S. The gas density associated with H$_2$S in this star is estimated to be around \(10^8\) cm\(^{-3}\), which is similar to the values for the galaxies in our study.
%\textcolor{red}{We can guess that H2S is tracing the gas different from the one which CO traces.}

In all these cases mentioned above, the sources are known to have high-velocity outflows and/or shocks associated. In the case of NGC\,253, \citet{Bouvier2024An253} conclude that there is strong evidence indicating that H$_2$S is tracing shocks. \textit{For the galactic sources M1-59 and W43A, they are known to exhibit high-velocity collimated outflows and shocks. Therefore, H$_2$S emission seems to be associated with shocks in a wide variety of astrophysical phenomena.}  
As mentioned above, our ALMA observations reveal the presence of relatively broad spectral line wings. This strongly suggests that the presence of H$_2$S in these galaxies could also be related to an outflow that compresses gas into compact high-density regions. Particularly, in the case of NGC\,1377, the enhancement of H$_2$S abundance could be associated with shocks, potentially created by the collimated outflow that has been observed in CO \citep[Fig.\ref{fig:1377CO}, and][]{Aalto2020ALMA1377}. In this scenario, CO would be tracing lower-density gas, potentially in an accretion disk and/or the base of a high-velocity outflow, whereas H$_2$S would be tracing a higher-density region.
 This region traced by H$_2$S could be also the densest parts of the gas in a clumpy torus around the nucleus.

It is important to note that while high-velocity gas is observed, the velocity experienced locally by the shocked gas may be lower than the bulk outflow speed. Indeed, molecules such as H$_2$S would be dissociated in the passage of fast shocks exceeding $\sim$50 km s$^{-1}$ \citep{Neufeld1989FASTSHOCK}. However, if sputtering induced by shocks is the dominant process releasing H$_2$S into the gas phase, as proposed by \citet{Bouvier2024An253}, this implies that gas and dust grains were already moving at significant velocities relative to their ambient medium before the shock impact. Consequently, the relative velocity between the gas and the shock front could be moderate enough to prevent complete molecular dissociation, thereby allowing H$_2$S to survive and be detected. In this scenario, H$_2$S traces denser regions where shocks act to liberate molecules without fully destroying them, even in environments with otherwise high-velocity outflows.
This interpretation aligns with the studies of NGC\,1266, where \citet{Pellegrini2013ShockGalaxy, Otter2024PullingGemini-NIFS} found that shocks are likely slow C-type ($\sim$30 km s$^{-1}$), consistent with conditions that preserve molecules like H$_2$S while still enabling their release from the ice mantles around dust grains.

%%%%%%%%%%%%%%%%%%%%%%%%%%%%%%%%%%%%%%%%%%%%%%%%%%%%%%%%%%%%%%

%%%%%%%%%%%%%%%%%%%%%%%%%%%%%%%%%%%%%%%%%%%%%%%%%%%%%%%%%%%%%%
\section{Conclusions}
\label{sec:conc}
%%%%%%%%%%%%%%%%%%%%%%%%%%%%%%%%%%%%%%%%%%%%%%%%%%%%%%%%%%%%%%
% The results indicate that H$_2$S emission is closely linked to the activity within very compact ($20-30$ pc) nuclear regions of these galaxies. The presence of high-velocity outflows and the derived high gas densities suggest that at least some component of this emission is related to shocks. \\
% - However, shocks alone cannot fully explain the observed abundance of gas-phase H$_2$S. Instead, a combination of processes likely contributes to its presence. \\
% - Consequently, H$_2$S should not be considered a direct tracer of shocks, as multiple mechanisms may enhance its abundance. \\
% - The association of H$_2$S with dense gas implies that shock compression may play a role in creating the necessary conditions for its excitation.\\
% - Further high-resolution observations, including more H$_2$S transitions and complementary shock tracers, will be needed to disentangle these processes and refine our understanding of the role of H$_2$S in molecular feedback and dense gas environments.

In this study, we have presented high-angular resolution ALMA observations of H$_2$S emission in a sample of nearby galaxies, providing new insights into the origin and excitation conditions of this molecule in extragalactic environments.

Our main conclusions are as follows:
\begin{itemize}
    \item H$_2$S emission is detected from centrally compact regions ($\lesssim$100–150 pc) in all observed galaxies, with some sources showing broad spectral line wings, indicative of kinematic components related to outflows or shocks. In NGC4418 H$_2$S also appears to be tracing gas that is counterrotating. A peculiar red-shifted emission feature may be inflowing gas, or possibly a slanted outflow.
    
    \item RADEX modelling suggests that the H$_2$S-emitting gas is characterised by high densities, with $n_{\rm H_2} \gtrsim 10^7$ cm$^{-3}$ in NGC\,4418 and NGC\,1266, and $n_{\rm H_2} > 3 \times 10^6$ cm$^{-3}$ in NGC\,1377. The derived densities are significantly higher than those inferred from CO observations, suggesting that H$_2$S traces a denser gas component.

    \item The high-density H$_2$S-emitting gas may originate from compact structures within the nuclear region, such as dense clumps, accretion flows onto a supermassive black hole, or shocked regions driven by molecular outflows.

    \item The presence of broad wings in the H$_2$S spectral profiles, combined with the detection of collimated CO outflows in some galaxies, supports a scenario where shocks associated with outflows contribute to the release of H$_2$S into the gas phase, likely via sputtering or thermal desorption from dust grains.

    \item Although high-velocity gas is present, the relative velocity of gas with respect to the shock front may be lower, preventing the complete dissociation of H$_2$S molecules even in fast outflows, as proposed in earlier studies \citep[e.g.,][]{Bouvier2024An253, Neufeld1989FASTSHOCK}.

    \item We also note important caveats related to the RADEX modelling, such as uncertainties introduced by assumptions about source size, the ortho-to-para ratio of H$_2$S, and the use of scaled collisional rates. These factors should be carefully considered when interpreting the derived physical conditions.

\end{itemize}

Overall, our results suggest that H$_2$S is a valuable tracer of dense, shocked gas in galactic nuclei and outflows, and further multi-line studies at higher spatial resolution will be crucial to fully characterise its role in feedback processes and molecular chemistry in active galaxies.

\begin{acknowledgements}
This paper makes use of the following ALMA data:
ADS/JAO.ALMA\#2018.1.00423.S, \#2018.1.01488.S, \#2018.1.00939.S, \#2012.1.00377.S, and \#2011.1.00511.S. ALMA is a partnership of ESO (representing its Member States), NSF (USA), and NINS (Japan), together with NRC(Canada) and NSC and ASIAA (Taiwan), in cooperation with the Republic of Chile. 
The Joint ALMA Observatory is operated by ESO, AUI/NRAO, and NAOJ. 
We thank the Nordic ALMA ARC node for excellent support.
The Nordic ARC node is funded through Swedish Research Council grant No 2017-00648. 
MS and SA gratefully acknowledge funding from
the European Research Council (ERC) under the European Union’s Horizon
2020 research and innovation programme (grant agreement No 789410, PI: S.
Aalto).
% S.A. gratefully acknowledges support from an ERC Advanced Grant 789410 and from the Swedish Research Council.
Part of this work was supported by the German\emph{Deut\-sche For\-schungs\-ge\-mein\-schaft, DFG\/} project number Ts~17/2--1.
S.V.  has received funding from the European Research Council (ERC) under the European Union’s Horizon 2020 research and innovation programme MOPPEX 833460.
\end{acknowledgements}

\bibliographystyle{aa} % style aa.bst
\bibliography{main.bib} % your references Yourfile.bib

\end{document}